\documentclass[journal]{IEEEtran}

\usepackage{cite}
\usepackage{graphicx}
\usepackage{amsmath}
\usepackage{algorithm}
\usepackage{amssymb}
\usepackage{diagbox}
\usepackage{epstopdf}
\usepackage{enumitem}
\usepackage[caption=false,font=footnotesize]{subfig}
\usepackage{threeparttable}
\usepackage{color}

\def\argmax{\text{argmax}}

\def\bC_k{\mathbb{C_k}}

\def\bE{\mathbb{E}}

\def\bR{\mathbb{R}}

\def\cC_k{\mathcal{C_k}}

\def\cI{\mathcal{I}}

\def\cN{\mathcal{N}}

\def \tilC_k{\widetilde{C_k}}

\def \haC_k{\widehat{C_k}}

\def \habfC_k{\widehat{\bf C_k}}

\def \bfC_k{{\bf C_k}}

\def\bee{\begin{equation}}
\def\ene{\end{equation}}
\def\beq{\begin{eqnarray}}
\def\enq{\end{eqnarray}}

\begin{document}
\title{Bayesian Filtering with Unknown Sensor Measurement Losses}

\author{Jiaqi~Zhang,
	Keyou~You,~\IEEEmembership{Senior Member,~IEEE,}
	and~Lihua~Xie,~\IEEEmembership{Fellow,~IEEE}% <-this % stops a space
	\thanks{J. Zhang and K. You are with the Department
		of Automation, and TNList, Tsinghua University, Beijing 100084, China. e-mail: zjq16@mails.tsinghua.edu.cn, youky@tsinghua.edu.cn.}% <-this % stops a space
	\thanks{L. Xie is with the School of Electrical and Electronic Engineering, Nanyang Technological University, Singapore 639798, Singapore. e-mail: elhxie@ntu.edu.sg.} }

\maketitle

\begin{abstract}
	This work studies the state estimation problem of a stochastic nonlinear system
	with {\em unknown} sensor measurement losses. If the estimator knows the sensor measurement losses of a linear Gaussian system, the minimum variance estimate is easily computed by the celebrated intermittent Kalman filter (IKF). However, this will no longer be the case when the measurement losses are unknown and/or the system is nonlinear or non-Gaussian. By exploiting the binary property of the measurement loss process and the IKF, we design three suboptimal filters for the state estimation, i.e., BKF-I, BKF-II and RBPF. The BKF-I is based on the MAP estimator of the measurement loss process and the BKF-II is derived by estimating the conditional loss probability.  The RBPF is a particle filter based algorithm which marginalizes out the loss process to increase the efficiency of particles. All the proposed filters can be easily implemented in recursive forms. Finally, a linear system, a target tracking system and a quadrotor's path control problem are included to illustrate their effectiveness, and show the tradeoff between computational complexity and estimation accuracy of the proposed filters.
\end{abstract}


% \fi
\IEEEpeerreviewmaketitle

\section{Introduction}
\IEEEPARstart{T}{he} Kalman filter (KF) has a simple structure in optimally correcting propagated state estimates with the sensor measurement of the observed system, and  has been successfully applied in countless guidance, navigation and control (GNC) related applications. While in some practical applications,  the estimator may not always access the true sensor measurement \cite{teo2010decentralized,soltani2011reliable,choukroun2012mode,yu2012hybrid,joerger2013kalman,avram2017nonlinear,zhang2009new,kottenstette2013design,reed2017jls}. For example, a sensor fault results in that the estimator only receives a pure noise \cite{yu2012hybrid,joerger2013kalman,avram2017nonlinear}, which does not contain any information of the estimated system.  In networked systems, the channel traffic congestion will also lead to sensor measurement losses \cite{zhang2009new}.  Similar issues commonly exist in the following important applications.
\begin{itemize}
	\item In a target tracking system, the position, distance, and velocity of the target to the base station is measured by a radar or Global Positioning System (GPS). The sensor information is sent to a missile or an unmanned aircraft via resource-limited wireless channels where communications between devices are power constrained and therefore limited in range and reliability \cite{zhang2009new,kottenstette2013design,reed2017jls}. If the true sensor measurement is lost, the base station only receives error data, e.g., channel noises.

	\item GPS is a common way for positioning, which however requires signals to be detected precisely. Thus any slight inaccuracy in the signal's reception or some external disturbances can lead to an error in the measured location \cite{soltani2011reliable}. Environmental factors such as trees, valleys, buildings, or even heavy cloud cover can impact or even interrupt the transmission of the GPS signal between the receiver and the satellites. Besides, urban environments surrounded by tall buildings also cause inaccurate GPS signals.

	\item A vehicle or aircraft is usually equipped with many sensors. For instance, a quadrotor may have an inertial measurement unit (IMU), a GPS receiver, barometers, and magnetometers. Large aircrafts even use more sensor units. It is possible that some error measurements occur from time to time due to unexpected faults in the system \cite{yu2012hybrid,choukroun2012mode,joerger2013kalman}.
\end{itemize}

In the above situations,  the KF or extended KF can only intermittently access the true sensor measurements. If an wrong measurement or pure noise is used for updating the state estimate,  the estimation performance will significantly degrade, and the filter may even diverge.

Then, a natural question is how to handle issues induced by occasional measurement errors or losses.  If the measurement loss is {\em known} to the estimator, the minimum variance estimate (MVE) of a linear Gaussian system is provided by an intermittent Kalman filter (IKF) \cite{sinopoli2004kalman}, and  a vast body of literature, see e.g. \cite{wu2017kalman,you2015analysis} and references therein, focuses on the stability of the IKF. For instance, it is shown in \cite{sinopoli2004kalman} that there exists a critical measurement loss probability beyond which the IKF may diverge. In \cite{you2011mean}, this critical value for certain type of systems is explicitly obtained under Markovian measurement losses. More cases can be found in \cite{rohr2014kalman}.

The problem is further complicated if the measurement loss is {\em unknown} to the estimator. In real applications, the estimator receives a data packet which may be a pure noise or fake sensor measurement. A faulty sensor might also return ``wrong" measurement data to the estimator. Thus, it is of practical importance to study the state estimation problem under {\em unknown} sensor measurement losses, which is the main focus of this work. Clearly, the unknown measurement loss results in a non-Gaussian and nonlinear system where the IKF is no longer applicable. Although there are some generic estimation methods for nonlinear and/or non-Gaussian systems, e.g., extended KF (EKF), unscented KF  \cite{van2000unscented,julier2004unscented} and particle filter (PF) \cite{arulampalam2002tutorial}, they do not particularly explore the unique feature of the current problem, in which the measurement loss or error results in a discontinuous measurement equation.  To resolve this issue,  we model the  process of sensor measurement losses as a binary sequence $\{\gamma_k\}$. Particularly, $\gamma_k=1$ means that the true sensor measurement is received and $\gamma_k=0$ indicates the occurrence of a sensor measurement loss. Our objective is to design practical suboptimal filters to accommodate the  unknown intermittent measurement losses in a nonlinear stochastic system. 

Motivated by the optimality of the IKF, our idea is that we first estimate the binary sequence $\{\gamma_k\}$ under the maximum a posteriori probability (MAP) estimation criterion. Following the principle of certainty equivalence, we then use the MAP estimate to replace the unknown measurement loss $\gamma_k$ in the IKF for the linearized systems. This results in our Bayesian Kalman filter I (BKF-I). We also derive a Bayesian Kalman filter II (BKF-II) by estimating the conditional measurement loss probability in the linearized system, which is a compromise between the KF and the IKF.  Clearly, both filters are very different from the existing nonlinear extended KF, unscented KF or PF, and  will reduce to the intermittent extended Kalman filter (IEKF) if the measurement loss process $\{\gamma_k\}$ is known to the estimator.

Another method to address the non-Gaussianity and/or non-linearity is the PF which samples particles to approximate the conditional density of state.
However, the amount of computations required for the PF in the high-dimensional state space is usually large. To increase the sampling efficiency, one can marginalize out some of the states and use the standard algorithms such as the KF to estimate them. Then, the PF is applied to estimate the rest of state variables, which is called Rao-Blackwellised particle filter (RBPF). The implementation and comparison between the standard PF and RBPF are well documented in \cite{doucet2001SMCintroduction,doucet2000rao,van2000unscented,gustafsson2002particle}. In this work, we adopt this idea by  using the PF to estimate the conditional distribution of the measurement loss process $\{\gamma_k\}$, which is a binary process and requires only a few number of particles to approximate its posterior distribution.  As in the BKF-I, the state of the linearized system is then estimated by the IKF. Compared to the standard PF, the efficiency of using particles is significantly improved, no matter how large the dimension of the system state is.

We compare the computational complexities and the estimation performance of the proposed suboptimal filters. Both the BKF-I and BKF-II have similar computational complexities to the IEKF. The computational complexity of the RBPF is essentially proportional to the number of particles. The larger number of particles is adopted in the RBPF, the better estimation accuracy is expected in general. Moreover, the RBPF weakly converges to an optimal MVE for a linear stochastic system with unknown measurement losses if the number of particles tends to infinity. Although the estimation accuracies of the BKF-I and BKF-II cannot be ensured, they are usually easier to implement. Thus, the choice of the three filters depends on the problem on hand.

Finally, we use the proposed filters to solve the state estimation problem of a linear stochastic system, a target tracking problem and a quadrotor's path control problem, all of which are subject to unknown measurement losses. In these examples, the proposed filters  well complete their estimation tasks under different levels of measurement losses, and their performance even comes close to the case with {\em known} measurement losses, showing that the measurement loss process $\{\gamma_k\}$ is correctly estimated.  For the RBPF, we also illustrate how the estimation accuracy is improved by increasing the number of particles in the linear system. This explains the tradeoff between computational complexity and the estimation accuracy of the RBPF.

It should be noted that the conference version of this paper \cite{zhang2016kalman} only studies linear stochastic systems. In comparison, this work focuses on nonlinear systems and provides detailed comparisons of the proposed three filters in terms of estimation accuracy and computational complexity, see Section \ref{sec_cmp} and Section \ref{sec_sim_exp1}. Besides, this paper includes the quadrotor's path control problem in Section \ref{sec_quadrotor} to validate the effectiveness of our filters and provides a fast algorithm for implementing the RBPF in Section \ref{sec_fast}. 

The rest of this paper is organized as follows. In Section \ref{sec_rcp}, we formulate the estimation problem with unknown measurement losses. In Section \ref{sec_bkf}, we design the BKF-I and the BKF-II  based on the Bayes' theorem and the technique of EKF. In Section \ref{rbpf}, we derive the RBPF to address the estimate of the process of unknown measurement losses. We compare the three filters both in terms of computational complexity and estimation accuracy in Section \ref{sec_cmp}.  Simulation is performed in Section \ref{sec_simulation} to show the effectiveness of the proposed filters and compare their performance. Finally, we draw some concluding remarks in Section \ref{sec_conclusion}.

\section{Problem formulation}
\label{sec_rcp}
In this section, we formulate the state estimation problem with unknown measurement losses and introduce the celebrated intermittent KF, whose optimality is ensured for a linear Gaussian system with known measurement losses.

\subsection{State Estimation with Unknown Measurement Losses}
Consider a discrete dynamical system with measurement losses and the additive noise of the form:
\bee
\begin{split}
	x_{k+1}&=f(x_{k},u_{k})+w_{k}\\
	y_{k}&=\vec{\gamma}_{k} \cdot h(x_{k})+v_{k}
\end{split}
\ene
where $x_k\in\bR^n$ and $y_k\in\bR^m$ are the vector state and measurement, respectively. The random vectors $w_k\in\bR^n$ and $v_k\in\bR^m$ are independent white Gaussian noises with zero means and covariance matrices $Q\geq0$ and $R\geq0$, respectively. The initial state $x_0$ is assumed to be a random Gaussian vector with mean $\bar{x}_0$ and covariance matrix $\Sigma_0>0$.
$\vec{\gamma}_{k}$ is a diagonal matrix with its $i$-th diagonal element $\gamma_{ik}, i\in\{1,2,.., m\}$, representing the $i$-th sensor measurement loss at time step $k$. In particular, $\gamma_{ik}=1$ indicates that the true sensor measurement $y_{ik}$ is successfully received by the estimator while $\gamma_{ik}=0$ means that the estimator only receives a fake measurement of $y_{ik}$, i.e., the pure noise $v_{ik}$.

In many real systems, all entries of $y_{k}$ are obtained from the same sensor and are transmitted as a single packet, e.g., the GPS signal usually consists of position and velocity measurements. Then, all $\gamma_{ik},i\in\{1,...,m\}$ are identical, and $\vec{\gamma}_{k}$ is reduced to a random variable. For brevity, this paper mainly studies this case and focuses on the following nonlinear stochastic system
\bee\label{nonlinear}
\begin{split}
	{x}_{k+1}&=f(x_{k},u_{k})+w_{k}\\
	y_{k}&=\gamma_{k}\cdot h(x_{k})+v_{k}
\end{split}
\ene
where  $\gamma_k$ is a binary random variable that represents the sensor measurement loss at time step $k$. Thus, $\gamma_k=1$ indicates that the true sensor measurement $y_k$ is contained in the arrival data packet while $\gamma_k=0$ means that the estimator only receives pure noise.

The goal of this work is to propose recursive filters to estimate the state of the system in  (\ref{nonlinear}) under unknown measurement loss process $\{\gamma_k\}$.

\subsection{Intermittent Extended KF (IEKF)}
For a linear Gaussian model, the IKF in \cite{sinopoli2004kalman} assumes that $\{\gamma_k\}$ is known to the estimator.
To revise it for the nonlinear system in (\ref{nonlinear}),  define $\Gamma_{k}=\{\gamma_0,...,\gamma_{k}\}$, $U_{k}=\{u_0,...,u_{k}\}$ and $Y_{k}=\{y_0,...,y_{k}\}$, and the conditional minimum variance estimate and error covariance matrices are given by
\bee\label{optimal}
\begin{split}
	\hat{x}_{k|k-1}&=\bE[x_k|Y_{k-1},\Gamma_{k-1},U_{k-1}]\\
	\hat{x}_{k|k}&=\bE[x_k|Y_k,\Gamma_{k},U_{k-1}]\\
	\Sigma_{k|k-1}&=\bE[(x_k-\hat{x}_{k|k-1})(x_k-\hat{x}_{k|k-1})^{\rm T}|Y_{k-1},\Gamma_{k-1},U_{k-1}]\\
	\Sigma_{k|k}&=\bE[(x_k-\hat{x}_{k|k})(x_k-\hat{x}_{k|k})^{\rm T}|Y_{k},\Gamma_{k},U_{k-1}]\\
	\hat{y}_{k|k-1}&=\bE[y_k|Y_{k-1},\Gamma_{k-1},U_{k-1}].\nonumber
\end{split}
\ene

Denote $A_k=\frac{\partial f(\hat{x}_{k|k},u_k)}{\partial x}$ and $C_k=\frac{\partial h(\hat{x}_{k|k-1})}{\partial x}$, this enables us to approximate (\ref{nonlinear}) by a linearized system
\bee\label{linearized}
\begin{split}
	x_{k+1}&=A_kx_k+w_k+b_k\\
	y_k&=\gamma_k\cdot C_kx_k+v_k+z_k
\end{split}
\ene
where $b_k$ and $z_k$ are computed online from the equations
$$b_k=f(\hat{x}_{k|k},u_k)-A_k\hat{x}_{k|k}, z_k=\gamma_k (h(\hat{x}_{k|k-1})-C_k \hat{x}_{k|k-1}).$$

If $\gamma_k$ is known to the estimator, the measurement update  for the nonlinear system (\ref{nonlinear}) can be obtained by applying the IKF \cite{sinopoli2004kalman} to the linearized system (\ref{linearized}), which leads to the Intermittent Extended KF (IEKF), i.e.,
\bee\label{KF}
\begin{split}
	\hat{x}_{k|k}&=\hat{x}_{k|k-1}+\gamma_kK_k(y_k-h(\hat{x}_{k|k-1}))\\
	\Sigma_{k|k}&=\Sigma_{k|k-1}-\gamma_kK_kC_k\Sigma_{k|k-1}
\end{split}
\ene
and the time update is the same as the EKF, i.e.,
\bee\label{KF2}
\begin{split}
	\hat{x}_{k+1|k}&=f(\hat{x}_{k|k}, u_k),\\
	\Sigma_{k+1|k}&=A_k\Sigma_{k|k}A_k^{\rm T}+Q,
\end{split}
\ene
where the Kalman gain $K_k=\Sigma_{k|k-1}C_k^{\rm T}(C_k\Sigma_{k|k-1}C_k^{\rm T}+R)^{-1}$ and $\hat{x}_{0|-1}=\bar{x}_0, \Sigma_{0|-1}=\Sigma_0$.

In the present situation, $\{\gamma_k\}$ is {\em unknown} to the estimator, which renders the above IEKF inapplicable. However, it is still very helpful in designing an effective filter in a recursive form.

%In this work, we propose three recursive suboptimal filters in the sequel for above problem. Since $U_{k}$ is always known to estimator, we assume $U_{k}=0$ for all $k$ in following sections without loss of generality.

\section{Bayesian Kalman Filters}
\label{sec_bkf}
In this section, we propose two suboptimal recursive filters to solve the filtering problem with unknown sensor measurement losses.

\subsection{Bayesian Kalman Filter I}
We design a nonlinear filter called Bayesian Kalman filter I (BKF-I) to recursively compute the state estimate with unknown measurement losses.  An intuitive idea, which is motivated by the principle of certainty equivalence, is that we first estimate the measurement losses $\Gamma_k$, based on which the IEKF \eqref{KF} is then applied to compute the state estimate. We shall elaborate it in this subsection.

Since $\gamma_k$ is binary, it is natural to adopt the maximum a posteriori probability (MAP) estimate, and the MAP estimate of $\Gamma_k$ is given as follows:
\bee
\begin{split}
	\widehat{\Gamma}_k=\argmax_{\Gamma_k}p(\Gamma_k|Y_k)
\end{split}\nonumber
\ene
where $\widehat{\Gamma}_{k}=\{\hat{\gamma}_0,...,\hat{\gamma}_{k}\}$ and for notional simplicity,  we directly use $p(x)$ to denote either the probability density or mass function of a random vector $X$. 

Substitute $\widehat{\Gamma}_k$ into \eqref{KF}, i.e., use the estimate $\hat{\gamma}_{k}$ to replace unknown measurement loss ${\gamma}_{k}$ in \eqref{KF}, we obtain the BKF-I, see Algorithm \ref{BKFI}. The remaining problem reduces to the derivation of the MAP estimate of $\Gamma_k$. To solve it, we use the Bayes' formulas and obtain that
\beq
p(\Gamma_k|Y_k)&=&p(\gamma_k,\Gamma_{k-1}|Y_k)\nonumber\\
&=&\frac{p(\gamma_k,\Gamma_{k-1},y_k|Y_{k-1})p(Y_{k-1})}{p(Y_k)}.
\nonumber
\enq
To recursively compute the above, we note that
\beq
&& p(\gamma_k,\Gamma_{k-1},y_k|Y_{k-1})\nonumber \\
&&=p(y_k|\gamma_k,\Gamma_{k-1},Y_{k-1})p(\gamma_k,\Gamma_{k-1}|Y_{k-1}) \nonumber \\
&&=p(y_k|\gamma_k,\Gamma_{k-1},Y_{k-1})p(\gamma_k|\Gamma_{k-1},Y_{k-1})p(\Gamma_{k-1}|Y_{k-1}).
\nonumber
\enq
This implies that
\bee\label{recursive}
\begin{split}
	&p(\Gamma_{k}|Y_k)=\frac{p(y_k|\gamma_k,\Gamma_{k-1},Y_{k-1})p(\gamma_k|\Gamma_{k-1},Y_{k-1})}{p(y_k|Y_{k-1})}\\
	&~~~~~~~~~~~~~~\times p(\Gamma_{k-1}|Y_{k-1}).
\end{split}
\ene
To compute $p(\Gamma_k|Y_k)$, it requires to consider all possible values of $\Gamma_k$, which grows unboundedly with respect to the time steps. Therefore, it is impossible to recursively obtain the MAP estimate of $\Gamma_k$. In real applications, recursive algorithms are essential. Thus, we consider to approximately compute $p(\Gamma_{k}|Y_k)$ in a recursive way. This is achieved by using the estimate $\widehat{\Gamma}_{k-1}$ rather than the unknown $\Gamma_{k-1}$ to estimate $\gamma_k$. 

By substituting $\Gamma_{k-1}$ with $\widehat{\Gamma}_{k-1}$, it follows from \eqref{recursive} that
\beq\label{appro1}
&&\hspace{-.5cm}p(\Gamma_k|Y_k)\approx p(\gamma_k,\widehat{\Gamma}_{k-1}|Y_k) \\
&&\propto p(y_k|\gamma_k,\widehat{\Gamma}_{k-1},Y_{k-1})p(\gamma_k|\widehat{\Gamma}_{k-1},Y_{k-1})p(\widehat{\Gamma}_{k-1}|Y_{k-1}).\nonumber
\enq

Since $\widehat{\Gamma}_{k-1}$ is already obtained at time step $k$, our objective is to find $\gamma_k$ to maximize the approximated posterior probability $p(\gamma_k,\widehat{\Gamma}_{k-1}|Y_k)$ in (\ref{appro1}). As $\gamma_k$ is a binary variable, this is easily solved by letting  
\beq\label{bayes}
&&\hspace{-0.5cm}\hat{\gamma}_k=\argmax_{\gamma_k}p(\gamma_k,\widehat{\Gamma}_{k-1}|Y_k)\\
&&=\left\{\begin{aligned}
	1 & , ~\text{if}~p(\gamma_k=1,\widehat{\Gamma}_{k-1}|Y_k)>p(\gamma_k=0,\widehat{\Gamma}_{k-1}|Y_k), \\
	0 & , ~\text{otherwise.}
\end{aligned}\right.\nonumber
\enq

In view of (\ref{appro1}), we further obtain that
\beq\label{bayes2}
&& \hspace{-.5cm} \frac{p(\gamma_k=1,\widehat{\Gamma}_{k-1}|Y_k)}{p(\gamma_k=0,\widehat{\Gamma}_{k-1}|Y_k)} \\
&&=\frac{p(y_k|\gamma_k=1,\widehat{\Gamma}_{k-1},Y_{k-1})p(\gamma_k=1|\widehat{\Gamma}_{k-1},Y_{k-1})}{p(y_k|\gamma_k=0,\widehat{\Gamma}_{k-1},Y_{k-1})p(\gamma_k=0|\widehat{\Gamma}_{k-1},Y_{k-1})}. \nonumber
\enq

Then, we compute $p(y_k|\gamma_k,\widehat{\Gamma}_{k-1},Y_{k-1})$ and $p(\gamma_k|\widehat{\Gamma}_{k-1},Y_{k-1})$.  To obtain $p(y_k|\gamma_k,\widehat{\Gamma}_{k-1},Y_{k-1})$, we denote the probability density function of the Gaussian distribution with mean $\mu$ and covariance matrix $\sigma^2$ by $\cN(\mu,\sigma^2)$. Once $\widehat{\Gamma}_k$ is known, it follows from the IEKF that
\beq
&&\hspace{-1cm}p(y_k|\gamma_k,Y_{k-1},\widehat{\Gamma}_{k-1})\\
&&\approx\left\{\begin{aligned}
	 & \cN(h(\hat{x}_{k|k-1}),C_k\Sigma_{k|k-1}C_k^{\rm T}+R), & \gamma_k=1, \\
	 & \cN(0,R),                                               & \gamma_k=0,
\end{aligned}\right. \nonumber
\enq
where $\hat{x}_{k|k-1}$ and $\Sigma_{k|k-1}$ are computed in \eqref{KF} by replacing $\Gamma_{k-1}$ with $\widehat{\Gamma}_{k-1}$.

If the prior probability distribution of $\gamma_k$ is also available, we are able to easily compute $p(\gamma_k|Y_{k-1},\widehat{\Gamma}_{k-1})$. Two common cases are illustrated below.
\begin{itemize}
	\item[(a)] $\{\gamma_k\}$ is a Bernoulli process with parameter $\theta$, then
	      \bee
	      p(\gamma_k=1|Y_{k-1},\widehat{\Gamma}_{k-1})=p(\gamma_k=1)=\theta.\nonumber
	      \ene
	\item[(b)] $\{\gamma_k\}$ is a Markov process with the transition probability matrix \cite{huang2007stability}
	      \bee
	      \centering p(\gamma_k=j|\gamma_{k-1}=i)=\begin{bmatrix} 1-q & q \\ p & 1-p \end{bmatrix}, i,j\in\{0,1\}\nonumber,
	      \ene
	      then $p(\gamma_k|Y_{k-1},\widehat{\Gamma}_{k-1})=p(\gamma_k|\hat{\gamma}_{k-1})$.
\end{itemize}

Finally, we can compute $p(\gamma_k,\widehat{\Gamma}_{k-1}|Y_k)$, and thus obtain the MAP estimate $\hat{\gamma}_k$ by (\ref{bayes}) and (\ref{bayes2}). The estimate $\hat{x}_{k|k}$ is then updated from \eqref{KF} by replacing $\gamma_k$ with $\hat{\gamma}_k$. 

Overall,  the BKF-I is summarized in Algorithm \ref{BKFI}. Compared with the IEKF, we need to further compute the MAP estimate of $\gamma_k$. The good news is that its computational complexity is still comparable to that of the IEKF.

\begin{algorithm}
	\caption{Bayesian Kalman filter I}\label{BKFI}
	\begin{enumerate}
		\item{\bf Initialization:}  Let $\hat{x}_{0|-1}=\bar{x}_0,\Sigma_{0|-1}=\Sigma_0$.
		\item{\bf MAP estimate of $\gamma_k$:} Given $y_k,\hat{x}_{k|k-1},\Sigma_{k|k-1}$ and $\widehat{\Gamma}_{k-1}$, compute
		  (\ref{bayes2}) and approximately obtain the MAP estimate of $\gamma_k$ by (\ref{bayes}).
		\item{\bf Measurement update:} Given $y_k$ and $\hat{\gamma}_k$, update the state estimate and its estimation error covariance matrix via (\ref{KF}) by replacing $\gamma_k$ with $\hat{\gamma}_k$.
		\item{\bf Time update:} Update the state prediction via (\ref{KF2}).
	\end{enumerate}
\end{algorithm}

\subsection{Bayesian Kalman Filter II}
In this subsection, we design the second suboptimal filter by approximately computing the posterior distribution of the measurement loss process $\{\gamma_k\}$.

Consider the minimum variance estimate, which is given by the conditional expectation
\bee
\hat{x}_{k|k}=\bE[x_k|Y_k]=\int x_kp(x_k|Y_k)dx_k.\nonumber
\ene

Since the system is nonlinear and non-Gaussian, the EKF cannot be applied to compute the optimal estimate. Since the posterior distribution $p(x_k|Y_k)$ is not tractable, the optimal filter is unavailable or cannot be implemented recursively. Note that a recursive suboptimal filter is necessary in practice. To this end, we adopt the following approximation
\bee
p(x_k|Y_k)\approx p(x_k|y_k,\hat{x}_{k|k-1}).\nonumber
\ene

That is, we use $\hat{x}_{k|k-1}$ to synthesize the information contained in $Y_{k-1}$ to compute the posterior distribution. Then, it follows that
\bee
\begin{split}
	\hat{x}_{k|k}&=\bE[x_k|Y_k]\approx \bE[x_k|y_k,\hat{x}_{k|k-1}],\\
	\Sigma_{k|k}&=\bE[(x_k-\bE(x_k|Y_k))(x_k-\bE(x_k|Y_k))^{\rm T}|Y_k]\\
	&\approx \bE[(x_k-\hat{x}_{k|k})(x_k-\hat{x}_{k|k})^{\rm T}|\hat{x}_{k|k-1},y_k].
\end{split}\nonumber
\ene

Besides, one can derive that
\bee
\begin{split}
	&p(x_k|\hat{x}_{k|k-1},y_k)\\
	&=\sum_{i=0}^{1}p(x_k|\hat{x}_{k|k-1},\gamma_k=i,y_k)p(\gamma_k=i|\hat{x}_{k|k-1},y_k)\\
	&=\lambda_k p(x_k|\gamma_k=1,\hat{x}_{k|k-1},y_k) \\
	&~~~~~+(1-\lambda_k)p(x_k|\gamma_k=0,\hat{x}_{k|k-1},y_k)
\end{split}\nonumber
\ene
where $\lambda_k$ denotes the probability of event $\{\gamma_k=1\}$ conditioned on $\hat{x}_{k|k-1}$ and $y_k$, and is given by
\bee\label{lambdak}
\begin{split}
	\lambda_k &=p(\gamma_k=1|\hat{x}_{k|k-1},y_k) \\
	&=\frac{p(y_k|\gamma_k=1,\hat{x}_{k|k-1})p(\gamma_k=1|\hat{x}_{k|k-1})}{\sum_{j=0}^{1}p(y_k|\gamma_k=j,\hat{x}_{k|k-1})p(\gamma_k=j|\hat{x}_{k|k-1})},\\
\end{split}
\ene
where $p(\gamma_k=1|\hat{x}_{k|k-1})$ is approximated by $\lambda_{k-1}$. 

Now we compute $p(y_k|\gamma_k,\hat{x}_{k|k-1})$. Consider a Gaussian approximation  $p(x_k|\hat{x}_{k|k-1})\approx \cN(\hat{x}_{k|k-1},\Sigma_{k|k-1})$, it yields that
\bee
\begin{split}
	&\hspace{-0.5cm}p(y_k|\gamma_k,\hat{x}_{k|k-1})\\
	&\approx \left\{\begin{aligned}
		 & \cN(h(\hat{x}_{k|k-1}),C_k\Sigma_{k|k-1}C_k^{\rm T}+R), & \gamma_k=1, \\
		 & \cN(0,R),                                               & \gamma_k=0.
	\end{aligned}\right.
\end{split}
\ene

Adopting the extended Kalman filtering technique \cite{anderson2012optimal}, we obtain that
\bee
\begin{split}
	p(x_k|\gamma_k,\hat{x}_{k|k-1},y_k)\approx\left\{\begin{aligned}
		 & \cN(\mu_1,\sigma_1^2), & \gamma_k=1, \\
		 & \cN(\mu_2,\sigma_2^2), & \gamma_k=0,
	\end{aligned}\right.
\end{split}\nonumber
\ene
where the mean and covariance are respectively given by
\bee\label{eq1_BKFII}
\begin{split}
	\mu_1&=\hat{x}_{k|k-1}+K_k(y_k-h(\hat{x}_{k|k-1})),\\
	\sigma_1^2&=\Sigma_{k|k-1}-\Sigma_{k|k-1}C_k^{\rm T}(C_k\Sigma_{k|k-1}C_k^{\rm T}+R)^{-1}C_k\Sigma_{k|k-1},\\
	\mu_2&=\hat{x}_{k|k-1},\\
	\sigma_2^2&=\Sigma_{k|k-1}.
\end{split}
\ene

Combining the above, the measurement update for the minimum variance estimate is approximately given as
\bee
\begin{split}
	\hat{x}_{k|k}&\approx\int x_kp(x_k|y_k,\hat{x}_{k|k-1})dx_k\\
	&=\lambda_k\int x_kp(x_k|\gamma_k=1,\hat{x}_{k|k-1},y_k)dx_k \\
	&~~~~~+(1-\lambda_k)\int x_kp(x_k|\gamma_k=0,\hat{x}_{k|k-1},y_k)dx_k\\
	&=\lambda_k\mu_1+(1-\lambda_k)\mu_2\\
	&=\hat{x}_{k|k-1}+\lambda_k K_k(y_k-h(\hat{x}_{k|k-1})).\\
\end{split}\nonumber
\ene
where the last equality follows from \eqref{eq1_BKFII}.

To run the algorithm recursively,  we still need to update  $\Sigma_{k|k}$ as well, which is derived below.
\bee
\begin{split}
	&\Sigma_{k|k}\\
	&\approx \bE[(x_k-\bE(x_k))(x_k-\bE(x_k))^{\rm T}|\hat{x}_{k|k-1},y_k]\\
	&=\bE[x_kx_k^{\rm T}|\hat{x}_{k|k-1},y_k]-\bE[x_k|\hat{x}_{k|k-1},y_k]\bE[x_k|\hat{x}_{k|k-1},y_k]^{\rm T}\\
	&\approx \bE[x_kx_k^{\rm T}|\hat{x}_{k|k-1},y_k]-\hat{x}_{k|k}\hat{x}_{k|k}^{\rm T}.
\end{split}\nonumber
\ene
The first term of $\Sigma_{k|k}$ is further written as
\bee
\begin{split}
	\bE&[x_kx_k^{\rm T}|\hat{x}_{k|k-1},y_k]=\int x_kx_k^{\rm T}p(x_k|\hat{x}_{k|k-1},y_k)dx_k\\
	&=\lambda_k\int x_kx_k^{\rm T}p(x_k|\gamma_k=1,\hat{x}_{k|k-1},y_k)dx_k+\\
	&~~~~~~(1-\lambda_k)\int x_kx_k^{\rm T}p(x_k|\gamma_k=0,\hat{x}_{k|k-1},y_k)dx_k\\
	&=\lambda_k\bE[x_kx_k^{\rm T}|\gamma_k=1,\hat{x}_{k|k-1},y_k]\\
	&~~~~~~+(1-\lambda_k)\bE[x_kx_k^{\rm T}|\gamma_k=0,\hat{x}_{k|k-1},y_k]\\
	&=\lambda_k(\mu_1\mu_1^{\rm T}+\sigma_1^2)+(1-\lambda_k)(\mu_2\mu_2^{\rm T}+\sigma_2^2).\\
\end{split}\nonumber
\ene
Finally, $\Sigma_{k|k}$ is updated by the following formula
\bee
\begin{split}
	&\Sigma_{k|k}\\
	&\approx\bE[x_kx_k^{\rm T}|\hat{x}_{k|k-1},y_k]-\hat{x}_{k|k}\hat{x}_{k|k}^{\rm T}\\
	&=\lambda_k(\mu_1\mu_1^{\rm T}+\sigma_1^2)+(1-\lambda_k)(\mu_2\mu_2^{\rm T}+\sigma_2^2)\\
	&~~~~~ -[\lambda_k\mu_1+(1-\lambda_k)\mu_2)][\lambda_k\mu_1+(1-\lambda_k)\mu_2)]^{\rm T}\\
	&=\lambda_k(1-\lambda_k)(\mu_1-\mu_2)(\mu_1-\mu_2)^{\rm T}+\lambda_k\sigma_1^2+(1-\lambda_k)\sigma_2^2\\
	&=\Sigma_{k|k-1}-\lambda_k K_kC_k\Sigma_{k|k-1}+\lambda_k(1-\lambda_k)\\
	&~~~~~\times [K_k(y_k-h(\hat{x}_{k|k-1}))(y_k-h(\hat{x}_{k|k-1}))^{\rm T}K_k^{\rm T}].\\
\end{split}\nonumber
\ene

Overall, the BKF-II is summarized in Algorithm \ref{BKFII}. In the BKF-II, we use $\lambda_k$ to represent the uncertainty induced by the measurement losses. By (\ref{lambdak}), it is clear that if $\gamma_k$ is known to the estimator, then $\lambda_k=\gamma_k$, which reduces to the IEKF. As $\gamma_k$ is {\em unknown} in our case,  $\lambda_k$ plays the role of estimating $\gamma_k$.

Note again that approximation is adopted to derive the recursive filter. However, from the storage and computation points of view, this suboptimal filter may be better than an optimal one in applications, since an optimal minimum variance estimate conditioned on $Y_k$ is generally not tractable.

\begin{algorithm}
	\caption{Bayesian Kalman filter II}\label{BKFII}
	\begin{enumerate}
		\item {\bf Initialization:}  Let $\hat{x}_{0|-1}=\bar{x}_0,\Sigma_{0|-1}=\Sigma_0$.
		\item {\bf Measurement update:} The state estimate and its approximated error covariance matrix are updated as follows.
		      \beq
		      \hat{x}_{k|k}\hspace{-.3cm}&=\hspace{-.3cm}&\hat{x}_{k|k-1}+\lambda_k K_k(y_k-h(\hat{x}_{k|k-1})),\nonumber\\
		      \Sigma_{k|k}\hspace{-.3cm}&=\hspace{-.3cm}&\Sigma_{k|k-1}-\lambda_k K_kC_k\Sigma_{k|k-1}+\lambda_k(1-\lambda_k)\nonumber\\
		      && [K_k(y_k-h(\hat{x}_{k|k-1}))(y_k-h(\hat{x}_{k|k-1}))^{\rm T}K_k^{\rm T}]\nonumber
		      \enq
		      where $\lambda_k$ is given in (\ref{lambdak}).
		\item {\bf Time update:} Update the state prediction via (\ref{KF2}).
	\end{enumerate}
\end{algorithm}

\section{Rao-Blackwellised Particle Filter}
\label{rbpf}
Clearly, both the BKF-I and BKF-II are not the minimum variance estimate and their performance cannot be guaranteed due to the use of approximation. In this section, a numerical method called Rao-Blackwellised particle filter (RBPF) is applied to approximately compute the minimum variance estimate. While the RBPF is computationally more demanding than the BKF-I and BKF-II, its estimation performance can be improved by increasing the number of particles.
\subsection{RB Particle Filter}
Recall that a minimum variance filter of $x_k$ is expressed by
\bee\label{condi}
\hat{x}_{k|k}=\int x_kp(x_k|Y_k)dx_k=\iint x_kp(x_k,\Gamma_k|Y_k)dx_kd\Gamma_k.
\ene

Since $p(x_k,\Gamma_k|Y_k)$ is not Gaussian, it is impossible to be analytically obtained, even for linear systems. Then, the integral is not computable and we have to resort to a numerical approach.

In this section, we choose a particle based algorithm to approximate the integral in (\ref{condi}). The particle filter (PF) is a powerful sampling based method to approximate any probability distribution, and address the non-linearity and non-Gaussianity problem. However, it is acknowledged that the number of particles grows dramatically with the dimension of the underlying random vector, which substantially increases the computation load.

In our case, if the particles are directly used to approximate $p(x_k,\Gamma_k|Y_k)$, it will result in a significant waste of particles since the unconditional distribution of the state for the linearized model in (\ref{linearized}) is deemed to be Gaussian, and the non-linearity and/or non-Gaussianity only appear in the measurement equation, which results from the unknown measurement losses. To well exploit this observation, we shall adopt the Rao-Blackwellised particle filter (RBPF) and use all the particles to approximate the conditional distribution of the binary variable $\gamma_k$. Intuitively, a few particles are sufficient to accomplish it. Furthermore, it follows from \cite{doucet2000rao} that the RBPF leads to better estimation performance than the standard PF.

To exposit it, it follows from the conditional probability definition that
\bee
p(x_k,\Gamma_k|Y_k)=p(x_k|\Gamma_k,Y_k)p(\Gamma_k|Y_k)\nonumber
\ene
where $p(x_k|\Gamma_k,Y_k)$ is an approximated Gaussian density and is recursively computed via the IEKF.

However, $p(\Gamma_k|Y_k)$ is difficult to obtain. As $\Gamma_k$ is a binary sequence, we use $N$ particles $\{\Gamma_k^i\}_{i=1}^N$ that are drawn from an importance density $q(\Gamma_k|Y_k)$ to approximate $p(\Gamma_k|Y_k)$, i.e.,
\bee\label{approxdis}
p(\Gamma_k|Y_k)\approx \sum_{i=1}^{N} {\omega}_{k}^i \delta(\Gamma_k-\Gamma_k^i),
\ene
where $ \delta(\cdot)$ is the standard Dirac delta function and ${\omega}_{k}^i$ is the normalized particle weight associated with $\Gamma_k^i$, i.e.,
\bee
{\omega}_{k}^i\propto\frac{p(\Gamma_k^i|Y_k)}{q(\Gamma_k^i|Y_k)}.\nonumber
\ene

Inserting (\ref{approxdis}) into (\ref{condi}), we obtain that
\bee\label{estimatpf}
\begin{split}
	\hat{x}_{k|k}&=\iint\!x_kp(x_k,\Gamma_k|Y_k)dx_kd\Gamma_k\\
	&=\iint\!x_kp(x_k|\Gamma_k,Y_k)dx_kp(\Gamma_k|Y_k)d\Gamma_k\\
	&\approx\sum_{i=1}^{N} {\omega}_{k}^i\int x_kp(x_k|\Gamma_k^i,Y_k)dx_k\\
	&=\sum_{i=1}^{N} {\omega}_{k}^i\bE[x_k|\Gamma_k^i,Y_k]=\sum_{i=1}^{N} {\omega}_{k}^i\hat{x}_{k|k}^i,
\end{split}
\ene
where $\hat{x}_{k|k}^i=\bE[x_k|\Gamma_k^i,Y_k]$.

Similarly, let $\Sigma_{k|k}^i=\bE[(x_k-\hat{x}_{k|k})(x_k-\hat{x}_{k|k})^{\rm T}|\Gamma_k^i,Y_k]$, then the estimation error covariance matrix is given as
\bee\label{errorcov}
\begin{aligned}
\Sigma_{k|k}&\approx\sum_{i=1}^{N} {\omega}_{k}^i\bE[(x_k-\hat{x}_{k|k})(x_k-\hat{x}_{k|k})^{\rm T}|\Gamma_k^i,Y_k]\\
&=\sum_{i=1}^{N} {\omega}_{k}^i\Sigma_{k|k}^i.
\end{aligned}
\ene

It should be noted from (\ref{optimal}) that both (\ref{estimatpf}) and (\ref{errorcov}) can be recursively computed by using the IEKF.
Specifically, we obtain that
% \bee
% \begin{split}
% 	&\bE[x_k|\Gamma_k^i,Y_k]=\hat{x}_{k|k-1}^i+\gamma_k^iK_k^i(y_k-h(\hat{x}_{k|k-1}^i)),\\
% 	&\bE[(x_k-\hat{x}_{k|k})(x_k-\hat{x}_{k|k})^{\rm T}|\Gamma_k^i,Y_k] \\
% 	&~~~~=\Sigma_{k|k-1}^i-\gamma_k^iK_k^iC_k\Sigma_{k|k-1}^i.
% \end{split}\nonumber
% \ene
\bee
\begin{split}
	&\hat{x}_{k|k}^i=\hat{x}_{k|k-1}^i+\gamma_k^iK_k^i(y_k-h(\hat{x}_{k|k-1}^i)),\\
	&\Sigma_{k|k}^i=\Sigma_{k|k-1}^i-\gamma_k^iK_k^iC_k\Sigma_{k|k-1}^i.
\end{split}\nonumber
\ene

The remaining problem is how to recursively generate particles $\{\Gamma_k^i\}$ and compute their associated weights $\{{\omega}_{k}^i\}$. This is resolved in next subsection.

\subsection{Importance Density}
If an importance density is chosen to factorize such that
\bee
q(\Gamma_k|Y_k)=q(\gamma_k|\Gamma_{k-1},Y_k)q(\Gamma_{k-1}|Y_{k-1}),\label{important}
\ene
one can obtain particles $\Gamma_k^i\sim q(\Gamma_k|Y_k)$ by augmenting each existing particle $\Gamma_{k-1}^i\sim q(\Gamma_{k-1}|Y_{k-1})$ with the new state $\gamma_k^i\sim q(\gamma_k|\Gamma_{k-1},Y_k)$ and  obtain a recursive filter algorithm. To elaborate it, we express $p(\Gamma_k|Y_k)$ in the following form
\bee
p(\Gamma_k|Y_k)\propto p(y_k|Y_{k-1},\Gamma_k)p(\gamma_k|\Gamma_{k-1},Y_{k-1})p(\Gamma_{k-1}|Y_{k-1}).\nonumber
\ene

Jointly with (\ref{important}), it implies that
\beq
w_k^i &\propto& \frac{p(y_k|Y_{k-1},\Gamma_k^i)p(\gamma_k^i|\Gamma_{k-1}^i,Y_{k-1})p(\Gamma_{k-1}^i|Y_{k-1})}{q(\gamma_k^i|\Gamma_{k-1}^i,Y_k)q(\Gamma_{k-1}^i|Y_{k-1})}\nonumber\\
&=&w_{k-1}^i \frac{p(y_k|Y_{k-1},\Gamma_k^i)p(\gamma_k^i|\Gamma_{k-1}^i,Y_{k-1})}{q(\gamma_k^i|\Gamma_{k-1}^i,Y_k)}.\label{weight}
\enq

A nice feature of the above is that $p(y_k|Y_{k-1},\Gamma_k^i)$ is approximately a conditional Gaussian density and is recursively computed by IEKF, i.e.,
\bee
\begin{split}
	&p(y_k|Y_{k-1},\Gamma_{k}^i)\\
	&\approx\left\{\begin{aligned}
		 & \cN(h(\hat{x}_{k|k-1}^i),C_k\Sigma_{k|k-1}^iC_k^{\rm T}+R), & \gamma_k^i=1, \\
		 & \cN(0,R),                                                   & \gamma_k^i=0.
	\end{aligned}\right.
\end{split}\nonumber
\ene

To alleviate the degeneracy problem, which is key to the success of PFs,  there are two good choices of the importance density $q(\gamma_k|\Gamma_{k-1}^i,Y_k)$ in the literature \cite{arulampalam2002tutorial}. One is
$p(\gamma_k|\Gamma_{k-1}^i,Y_k)$, which minimizes a suitable measure of the degeneracy of the algorithm, i.e.,
\bee
N_{\text{eff}}=\frac{N}{1+\text{var}(w_k^{*i})}\nonumber
\ene
where $w_k^{*i}=p(\gamma_k^i|Y_k)/q(\gamma_k|\Gamma_{k-1}^i,Y_k)$ is referred to as the true importance weights.

The other is $p(\gamma_k|\gamma_{k-1}^i)$, which makes it easy to draw particles and compute the important weights. Since $p(\gamma_k|\Gamma_{k-1},Y_k)$ is usually difficult to access, we choose $q(\gamma_k|\Gamma_{k-1},Y_k)=p(\gamma_k|\gamma_{k-1})$ as the importance density in this work. That is, the new particle is generated from the following distribution
\bee
\gamma_k^i\sim p(\gamma_k|\gamma_{k-1}^i).\nonumber
\ene

% \begin{rem}
If $\{\gamma_k\}$ is a Bernoulli process or Markov process, the update equation for the particle weight in (\ref{weight}) becomes particularly simple. Specifically,
\bee
\begin{split}
	\omega_{k}^i&\propto \frac{p(\Gamma_k^i|Y_k)}{q(\Gamma_k^i|Y_k)}=\omega_{k-1}^i\frac{p(y_k|Y_{k-1},\Gamma_k^i)p(\gamma_k^i|\gamma_{k-1}^i)}{q(\gamma_k^i|\Gamma_{k-1}^i,Y_k)}\\
	&=\omega_{k-1}^ip(y_k|Y_{k-1},\Gamma_k^i).
\end{split}\nonumber
\ene
% \end{rem}

In the current situation, the number of particles is very small since  $\gamma_k$  is binary. The RBPF with a resampling step is summarized in Algorithm \ref{al_pf}.

\begin{algorithm}\caption{Rao-Blackwellised particle filter}\label{al_pf}

	\begin{enumerate}
		\item {\bf Initialization:}  For $i=1,...,N.$ draw $N$ particles $\gamma_0^i$ from the prior $p(\gamma_0)$ and let $\hat{x}_{-1|0}^i=x_0,\Sigma_{-1|0}^i=\Sigma_0$.
		\item {\bf Importance sampling:}
		      \begin{enumerate}
			      \item For $i\in\{1,...,N\}$ and $k>0$, draw $N$ particles from the importance distribution $\gamma_k^i\sim p(\gamma_k|\gamma_{k-1}^i)$.
			      \item For $i\in\{1,...,N\}$ and given $y_k$, update the normalized importance weights $\omega_{k}^i\propto\omega_{k-1}^ip(y_k|Y_{k-1},\Gamma_k^i).$
		      \end{enumerate}

		\item {\bf Resampling:}
		      \begin{enumerate}
			      \item Compute an estimate of the effective number of particles $$\widehat{N}_{\text{eff}}=\frac{1}{\sum_{i=1}^{N}({\omega}_{k}^i)^2}.$$
			      \item  If $\widehat{N}_{\text{eff}}<N_T$, which is a prescribed threshold, perform resampling.
			            Take $N$ new samples $(\gamma_k^{i*},\hat{x}_{k|k-1}^{i*},\Sigma_{k|k-1}^{i*})$ with replacement
			            from the list $(\gamma_k^{i},\hat{x}_{k|k-1}^{i},\Sigma_{k|k-1}^{i})$, $i\in\{1,\ldots,N\}$ according to the probability distribution that $\text{Pr}\{\gamma_k^{i*}=\gamma_k^i\}={\omega}_{k}^i.$
			      \item Let $(\gamma_k^{i},\hat{x}_{k|k-1}^{i},\Sigma_{k|k-1}^{i})=(\gamma_k^{i*},\hat{x}_{k|k-1}^{i*},\Sigma_{k|k-1}^{i*})$ and  ${\omega}_{k}^i=1/N$ for all $i\in\{1,\ldots, N\}$.
		      \end{enumerate}
		\item {\bf Output}
		      \begin{enumerate}
			      \item  For each triple $(\gamma_k^{i},\hat{x}_{k|k-1}^{i},\Sigma_{k|k-1}^{i})$ and $y_k$, do
			            \bee
			            \begin{split}
				            K_k^i&=\Sigma_{k|k-1}^iC_k^{\rm T}(C_k\Sigma_{k|k-1}^iC_k^{\rm T}+R)^{-1}\\
				            \hat{x}_{k|k}^i&=\hat{x}_{k|k-1}^i+\gamma_k^iK_k^i(y_k-h(\hat{x}_{k|k-1}^i))\\
				            \Sigma_{k|k}^i&=\Sigma_{k|k-1}^i-\gamma_k^iK_kC_k\Sigma_{k|k-1}^i.
			            \end{split}\nonumber
			            \ene
			      \item Measurement update
			            \beq
			            \hat{x}_{k|k}&=\sum_{i=1}^{N} {\omega}_{k}^i\hat{x}_{k|k}^i,\  \Sigma_{k|k}=\sum_{i=1}^{N} {\omega}_{k}^i\Sigma_{k|k}^i.\nonumber
			            \enq
			      \item Update the state prediction by (\ref{KF2}), and set
			            \bee
			            \hspace{-0.3cm}\hat{x}_{k+1|k}^i=f(\hat{x}_{k|k}^i,u_k),\  \Sigma_{k+1|k}^i=A_k\Sigma_{k|k}^iA_k^\text{T}+Q.\nonumber
			            \ene
			          		  \end{enumerate}
					  \item   {\bf Let} $k=k+1$.

	\end{enumerate}
\end{algorithm}
\subsection{Fast Implementation}
\label{sec_fast}
The major computation of the RBPF is taken in the {\bf Output} step in Algorithm \ref{al_pf}, which requires an IEKF iteration for each triple $(\gamma_k^{i},\hat{x}_{k|k-1}^{i},\Sigma_{k|k-1}^{i})$. Thus, the computational complexity essentially is  proportional to the number of particles. When the number of particles is very large, one may observe that there are many duplicate triples after the {\bf Resampling}  c), i.e., there may be only $m\ll N$ triples that take different values. We can take advantage of this observation to reduce the computational complexity of RBPF. 

Instead of directly executing  an IEKF iteration for each triple $(\gamma_k^{i},\hat{x}_{k|k-1}^{i},\Sigma_{k|k-1}^{i})$, we first remove the duplicate triples from the list and obtain a set of triples that take different values, see step a) in Algorithm \ref{al_frbpf}, and then do an IEKF iteration for each of them, see step b) in Algorithm \ref{al_frbpf}. Finally, we set the updated value for the whole list, see step c). 

Clearly, Algorithm \ref{al_frbpf} returns the same estimate and prediction as those of RBPF in  Algorithm \ref{al_pf}. It may significantly reduce the computational complexity of the RBPF, especially when the number of particles is large. Besides, the  reduction of the computational complexity reflects the effectiveness of particles to some extent. However, when the number of particles is small, the fast implementation may be not efficient as it further requires to remove duplicate triples. This will be illustrated in Section  \ref{sec_sim_exp1} by an numerical example.

\begin{algorithm}\caption{Fast RBPF}\label{al_frbpf}
The {\bf Output} step in Algorithm \ref{al_pf} is replaced as follows. 
		      \begin{enumerate}
		      \renewcommand{\labelenumi}{\rm \alph{enumi})}
			      \item For the list of triples $(\gamma_k^{j},\hat{x}_{k|k-1}^{j},\Sigma_{k|k-1}^{j})$, $j\in\{1,...,N\}$, remove duplicate triples and obtain an index set $\cI$ such that there is no duplicate triple in the set $\{(\gamma_k^{i},\hat{x}_{k|k-1}^{i},\Sigma_{k|k-1}^{i}),i\in\cI\}$.
			      \item  For each triple $(\gamma_k^{i},\hat{x}_{k|k-1}^{i},\Sigma_{k|k-1}^{i}),i\in\cI$ and $y_k$, do
			            \bee
			            \begin{split}
				            K_k^i&=\Sigma_{k|k-1}^iC_k^{\rm T}(C_k\Sigma_{k|k-1}^iC_k^{\rm T}+R)^{-1}\\
				            \hat{x}_{k|k}^i&=\hat{x}_{k|k-1}^i+\gamma_k^iK_k^i(y_k-h(\hat{x}_{k|k-1}^i))\\
				            \Sigma_{k|k}^i&=\Sigma_{k|k-1}^i-\gamma_k^iK_kC_k\Sigma_{k|k-1}^i\nonumber\\
			            \hat{x}_{k+1|k}^i&=f(\hat{x}_{k|k}^i,u_k),\nonumber\\
			              \Sigma_{k+1|k}^i&=A_k\Sigma_{k|k}^iA_k^\text{T}+Q.\nonumber
			             \end{split}
			            \ene
			      \item For each $j\in\{1,...,N\}$, set
			            \[
				            \begin{aligned}
					            (\hat{x}_{k+1|k}^{j},\ \hat{x}_{k|k}^{j},\  & \hat{\Sigma}_{k+1|k}^{j},\ \hat{\Sigma}_{k|k}^{j}) \\=&(\hat{x}_{k+1|k}^{i},\ \hat{x}_{k|k}^{i},\ \hat{\Sigma}_{k+1|k}^{i},\ \hat{\Sigma}_{k|k}^{i})
				            \end{aligned}\] if $(\gamma_k^{j},\hat{x}_{k|k-1}^{j},\Sigma_{k|k-1}^{j})=(\gamma_k^{i},\hat{x}_{k|k-1}^{i},\Sigma_{k|k-1}^{i})$ for some $i\in\cI$.
			      \item Do measurement update
			            \beq
			            \hat{x}_{k|k}=\sum_{i=1}^{N} {\omega}_{k}^i\hat{x}_{k|k}^i,\  \Sigma_{k|k}=\sum_{i=1}^{N} {\omega}_{k}^i\Sigma_{k|k}^i,\nonumber
			            \enq
			            and update the state prediction by (\ref{KF2}).
		      \end{enumerate}
\end{algorithm}
\section{Comparisons of the three filters}\label{sec_cmp}
This section provides some comparisons of the proposed filters in terms of the estimation accuracy and computational complexity. We shall further verify the major results of this section by numerical examples in the next section.

\subsection{Estimation Accuracy}

If both $f$ and $h$ in (\ref{nonlinear}) are linear, any filter cannot outperform the optimal IKF, which however requires $\{\gamma_{k}\}$ is known. The performance of the RBPF can approach that of the IKF arbitrarily well by increasing the number of particles, and hence has a predictable performance. This cannot be guaranteed for the BKF-I and BKF-II due to approximations in deriving the recursive filters. However, if $y_k-h(\hat{x}_{k|k-1})$ is very different from $0$, the measurement loss $\gamma_{k}$ can be estimated correctly with high probability. Then, both the BKF-I and BKF-II are likely to approach the IKF. Compared to the BKF-II, the BKF-I is more intuitive and easier to implement. The BKF-II appears to be more stable as $\lambda_k$ retains the probability information of $\{\gamma_{k}=1\}$. Since the system in (\ref{nonlinear}) is nonlinear under the unknown measurement losses, these observations are impossible to prove in theory. We shall provide numerical examples for validation in the next section.

If either $f$ or $h$ in (\ref{nonlinear}) is nonlinear, our filters first linearize the nonlinear function and then applies the corresponding algorithms to the system in (\ref{linearized}). Due to the approximation in the linearization step, the IEKF is not optimal even it uses the true $\Gamma_{k}$, and all our filters might have better performance than the IEKF. Note that even when there is no measurement loss, i.e., $\gamma_k=1$ for all $k$, the estimation accuracy of the EKF cannot be guaranteed. {\color{black}By using a sufficiently large number of particles, the estimation accuracy of the RBPF may  even outperform that of the IEKF}.  Since approximation is used twice in both the BKF-I and BKF-II, it is expected that the IEKF and RBPF usually outperform the BKF-I and BKF-II. 
\subsection{Computational Complexity}
We provides Table \ref{table_complexity} to compare the computational complexities of the three proposed filters. The computation for linearization in deriving the system in (\ref{linearized}) is omitted since all the proposed filters require it.

The computational complexity can be quantified by the number of times of evaluating the Gaussian probability density function (PDF) and the matrix multiplications in one iteration. It is apparent that the RBPF has the highest computational complexity which is proportional to the number of particles and is far more than that of BKF-I and BKF-II. When $N$ is large, the fast RBPF might significantly reduce the computational complexity, and will be illustrated by numerical examples in the next section.

The BKF-I and BKF-II have similar computational complexities, and both require additionally two Gaussian PDF evaluations than that of the IEKF. Generally speaking, the computational complexity of evaluating a Gaussian PDF is $O(n^3)$, where $n$ is the dimension of the random vector and is the dimension of $x_k$ here.  Since the computational complexity of multiplication of two $n\times n$ matrices is also $O(n^3)$ in general, both the BKF-I and BKF-II have the same computational complexity $O(n^3)$, and are similar to that of the IEKF and EKF.
% \begin{table}[!t]\renewcommand{\arraystretch}{1.4}
% 	\centering
% 	\caption{Computational complexities in one iteration ($N$ is the number of particles in RBPF, and $n$ is the dimension of state vectors)}\label{table_complexity}
% 	\begin{tabular}{|c|c|c|c|c|}
% 		\hline Filters                          & IEKF     & RBPF             & BKF-I    & BKF-II   \\
% 		\hline       \begin{tabular}[c]{@{}c@{}} Number of times of\\ Gaussian PDF evaluations\end{tabular} & 0        & $2N$             & 2        & 2        \\
% 		\hline  \begin{tabular}[c]{@{}c@{}} Number of times of\\ matrix multiplications\end{tabular}      & 7        & \emph{$O(N)$}    & 7        & 9        \\
% 		\hline  Computational complexity           & $O(n^3)$ & \emph{$O(Nn^3)$} & $O(n^3)$ & $O(n^3)$ \\
% 		\hline
% 	\end{tabular}
% \end{table}

\begin{table}[!t]\renewcommand{\arraystretch}{1.4}
	\centering
	\begin{threeparttable}[b]
	\caption{Computational complexities in one iteration}\label{table_complexity}
	\begin{tabular}{|c|c|c|c|c|}
		\hline Filters                          & IEKF     & RBPF             & BKF-I    & BKF-II   \\
		\hline       \begin{tabular}[c]{@{}c@{}} Number of times of\\ Gaussian PDF evaluations\end{tabular} & 0        & $2N$             & 2        & 2        \\
		\hline  \begin{tabular}[c]{@{}c@{}} Number of times of\\ matrix multiplications\end{tabular}      & 7        & \emph{$O(N)$}    & 7        & 9        \\
		\hline  Computational complexity           & $O(n^3)$ & \emph{$O(Nn^3)$} & $O(n^3)$ & $O(n^3)$ \\
		\hline
	\end{tabular}
	\begin{tablenotes}
	\item [a] $N$ is the number of particles in RBPF.
	\item [b] $n$ is the dimension of state vectors.
	\end{tablenotes}	
	\end{threeparttable}
\end{table}

\subsection{Filter Selection}
The filter selection depends on many factors, and usually requires to find a good tradeoff between the estimation accuracy and the computational complexity. The tradeoff between the estimation accuracy and the computational complexity is illustrated in Section \ref{sec_sim_exp1}.
 If the computation resource is abundant, it is recommended to use the RBPF with a sufficient number of particles to obtain a good estimation accuracy.  If not, we may consider to use the BKF-I or BKF-II. Note that whether one filter is always superior to the other two filters is not conclusive. 

%We briefly summarize our previous conclusions on choice between the proposed filters.
%
%For a linear original system, all filters can achieve satisfactory performance. The RBPF with enough particles is more stable, but slower. The Fast RBPF can significantly reduce the computation time when the number of particles is large, while achieve the same performance. The BKF-I and BKF-II are easier to implement and fast.
%
%For a nonlinear original system, the performance of the three filters depends on the specific problem. We can also expect the RBPF with enough particles to have a better and stable performance in general, but BKF-I and BKF-II may be outperform it sometimes.

\section{Application Examples}
\label{sec_simulation}
In this section we adopt the proposed filters to solve three estimation problems. The first one is a state estimation problem of a linear system with randomly unknown measurement losses. In this example, it does not need to do linearization as in (\ref{linearized}). Then, the approximation in both BKF-I and BKF-II is solely used to handle the measurement loss process $\{\gamma_k\}$ in a recursive way. We compare the performance of proposed filters and show the relations between the number of particles, the computation time, and the root-mean-square error (RMSE) of the RBPF. 

The second one is a target tracking problem, which is a nonlinear system with measurement losses, and illustrates the effectiveness of our filters. We also provide a case where our filters do not work very well and give some explanations of it in this example. The last example is a quadrotor's path control problem with random measurement losses, where the quadrotor's states estimated by our filters are used in feedback to control a quadrotor. This requires precise state estimates, otherwise the system may diverge quickly. The result shows the quadrotor can complete the task well under all our filters.

\subsection{Linear System with Random Measurement Losses}\label{sec_sim_exp1}
Consider the following system
\bee\label{eq1_exp1}
\begin{aligned}
	x_{k+1} & =\begin{bmatrix}
		0.6 & 0.4 \\
		0.1 & 0.9
	\end{bmatrix}x_{k}+w_k,        \\
	y_{k}   & =\gamma_{k}\begin{bmatrix}
		1 & -2
	\end{bmatrix}x_k+v_k,
\end{aligned}
\ene
where $w_k$ and $v_k$ are independent Gaussian noise with zero means, and the covariance matrices are both identity matrices. $\{\gamma_{k}\}$ is an i.i.d. Bernoulli process with $Pr\{{\gamma_{k}=0}\}=p$. The initial state is also a Gaussian random vector with zero mean and identity covariance matrix. Since the open-loop poles of the system (\ref{eq1_exp1}) are $\lambda_1=0.5$ and $\lambda_2=1$, it follows from \cite{you2011mean} that the optimal IKF is expected to be mean square stable if the measurement loss level $p$ is strictly less than one.  

\begin{figure}[!t]
	\centering
	\includegraphics[width=\linewidth]{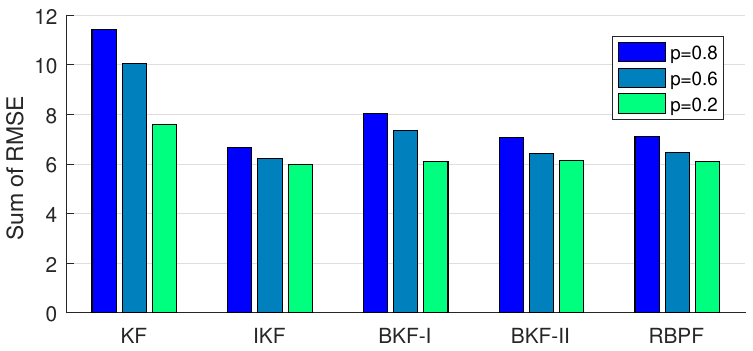}
	\caption{The sum of RMSE over all the $200$ time steps under different levels of measurement losses.}
	\label{fig1:exp3}
\end{figure}

To validate the proposed filters, we use the same sequence of noisy observations $Y_{200}$ for each filter, and the RBPF is implemented by using 20 particles. We also compare them with the optimal IKF, which requires to know the measurement loss, and the filter that does not consider the measurement loss, i.e.,  we simply set $\gamma_{k}=1$ for all $k$ in \eqref{KF} and \eqref{KF2},  which is denoted as `KF' in Fig. \ref{fig1:exp3}. 
To compute the RMSE of each estimate where the RMSE of an estimate $\hat{x}_{k|k}$ at time $k$ is defined as $(E[\|x_k-\hat{x}_{k|k}\|^2])^{1/2}$, we adopt the Monte Carlo method with $500$ independent experiments. Fig. \ref{fig1:exp3} illustrates that the estimation accuracy of the proposed three filters comes close to that of the optimal IKF, and is much better than that of the `KF'.  Moreover, the larger probability of the measurement losses, the better of the  improvement of the estimation accuracy over the `KF'. This clearly shows the advantage of considering the measurement losses in the filter design.

\begin{figure}[!t]
	\centering
	\includegraphics[width=\linewidth]{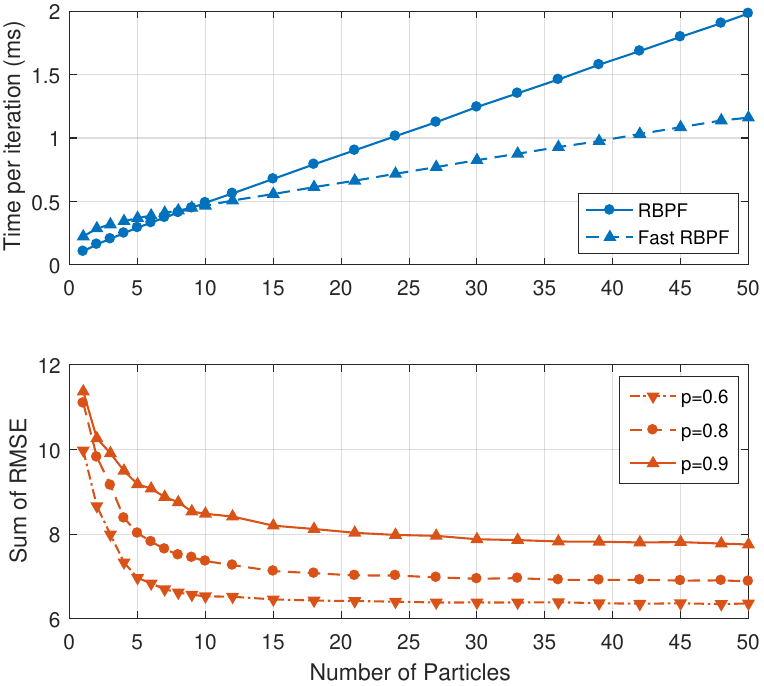}
	\caption{Computation time per iteration and RMSE of the RBPF with respect to the number of particles. The time is measured using Matlab.}
	\label{fig2:exp3}
\end{figure}

In Fig. \ref{fig2:exp3}, we illustrate how the number of particles affects the computational complexity and the estimation accuracy of the RBPF. One can easily observe that the computational complexity per iteration of the RBPF is proportional to the number of particles, which validates the result for the RBPF in Table \ref{table_complexity}. It also verifies that the fast implementation of RBPF in Algorithm \ref{al_frbpf} can greatly reduce the computation load when $N$ is large, while it is slightly slower than the RBPF for a small number of particles. The reason is that the fast RBPF further requires `find and set' steps, see steps a) and c) in Algorithm \ref{al_frbpf}, which are dominated by the computation for executing the IEKF iteration if the number of particles is large. Moreover, Fig. \ref{fig2:exp3} reveals that the sum of RMSE reduces quickly when $N$ is smaller, and slowly when $N$ is large. This suggests the existence of a critical number of particles that is vital to the tradeoff between the estimation accuracy and the computational complexity.  Note that the critical number is also related to the levels of random measurement losses.

\subsection{Target Tracking}
In this example, we test our filters on a  target tracking problem, which further requires the linearization in (\ref{linearized}). 
Consider a 2D target tracking problem where a radar is positioned on the ground to measure the distance and the angle of a moving target to the base station. The radar measurements are then sent to an unmanned tracking aircraft via a lossy network, which is subject to unknown measurement losses.

The motion and measurement equations are described by
\bee
\begin{split}
	x_{k+1}^{(i)}&=Fx_{k}^{(i)}+u_k^{(i)}+w_{k}^{(i)},i\in\{1, 2\}\\
	y_k&=\gamma_k\begin{bmatrix} r\\\phi \end{bmatrix}+ \begin{bmatrix} v_k^r\\v_k^\phi \end{bmatrix}\\
\end{split}\label{tracking}
\ene
where $x_k^{(i)}=[p_k^{(i)},\dot{p}_k^{(i)},\ddot{p}_k^{(i)}]^\text{T}$ denotes the target state at
time step $k$, including the target position, speed and acceleration in the $i$-th dimension. The system input $u_k^{(i)}$ is given as a prior, and the system matrices are
\bee
% \begin{split}
F= \begin{bmatrix} 1 & \tau & \tau^2/2 \\ 0 & 1 & \tau \\ 0 & 0 & 1 \end{bmatrix},%,G=\begin{bmatrix} T^3/6 \\ T^2/2 \\ T \end{bmatrix}\\
\begin{bmatrix} r\\\phi \end{bmatrix} =\begin{bmatrix} \sqrt{(p_k^{(1)})^2+(p_k^{(2)})^2} \\ \arctan\frac{p_k^{(2)}}{p_k^{(1)}} \end{bmatrix},\nonumber
% \end{split}
\ene
%\end{gather*}
where $\tau$ is the sampling period.

Here the measurement loss process $\{\gamma_k\}$ is also a Bernoulli process, which represents the loss of the radar  measurement. The input noise $w_k^{(i)}$ and the output noise $v_k=[v_k^r,v_k^\phi]^\text{T}$ are independently Gaussian distributed with zero mean.  As in  \cite{singer1970estimating} and \cite{gustafsson2002particle},  the covariance matrix of $w_k$ is set to
\bee
Q=2\alpha\sigma_m^2
\begin{bmatrix} \tau^5/20 & \tau^4/8 & \tau^3/6 \\ \tau^4/8 & \tau^3/3 & \tau^2/2 \\ \tau^3/6 & \tau^2/2 & \tau \end{bmatrix}\\\nonumber
\ene
and the covariance matrix of $v_k$ is given as
\bee
R=
\begin{bmatrix} \sigma_r^2 & 0 \\ 0 & \sigma_\phi^2\end{bmatrix}.\nonumber
\ene

The initial state $x_0^{(i)}$ is a Gaussian
random vector with mean $\bar{x}_0^{(i)}=[10, 0, 0]^\text{T},\ i\in\{1,2\}$ and covariance
\bee
\Sigma_0=
\begin{bmatrix} \sigma_r^2 & \sigma_r^2/\tau & 0 \\ \sigma_r^2/\tau & 2\sigma_r^2/\tau^2 & 0 \\ 0 & 0 & 0 \end{bmatrix}.\nonumber
\ene
\begin{table}[!t]
	\centering
	\caption{Parameters in the target tracking system}\label{tab:exp1}
	\renewcommand{\arraystretch}{1.3}
	\begin{tabular}{|c|c|c|}
		\hline
		Parameter     & Description                       & Value                  \\
		\hline
		$\tau$           & Sampling period                   & 0.01s                  \\
		$\sigma_m$    & Variance of target acceleration   & $4\text{m}/\text{s}^2$ \\
		$\alpha$      & \begin{tabular}[c]{@{}c@{}}Reciprocal of the\\ maneuver (acceleration) time constant\end{tabular}        & $1$                    \\
		$\sigma_\phi$ & \begin{tabular}[c]{@{}c@{}}Standard deviation of\\ angular measurement noise\end{tabular}        & $5^\circ$              \\
		$\sigma_r$    & \begin{tabular}[c]{@{}c@{}}Standard deviation of\\ distance measurement noise\end{tabular}        & 5m                     \\
		$p$           & Probability of measurement losses & 0.1, 0.3, 0.5, 0.7     \\
		\hline
	\end{tabular}
\end{table}

As in (\ref{linearized}), we obtain the linearized measurement equation
\bee
\renewcommand\arraystretch{2}
y_k=\gamma_k\begin{bmatrix} \frac{p_{k|k-1}^{(1)}}{r_{k|k-1}}    & \frac{p_{k|k-1}^{(2)}}{r_{k|k-1}}    \\
	\frac{-p_{k|k-1}^{(2)}}{r_{k|k-1}^2} & \frac{-p_{k|k-1}^{(1)}}{r_{k|k-1}^2}\end{bmatrix}\begin{bmatrix} p_k^{(1)}\\ p_k^{(2)}\end{bmatrix} +v_k+z_k, \nonumber
\ene
where $z_k$ is a function of $p_{k|k-1}$. All the parameters are of the same as those in \cite{singer1970estimating,curry2005radar}, see Table \ref{tab:exp1}, and the initial estimate is set to $\hat{x}_{0|0}^{(i)}=[10, 0, 0]^\text{T},\ i\in\{1,2\}$.

%\begin{table}[!t]
%\centering
%\caption{RMSE of the estimated position}\label{table}
%\renewcommand{\arraystretch}{1.0}
% \begin{tabular}{|c|c|c|c|c|}
% \hline
% \diagbox[trim=l]{Error rate}{ALG} &RBPF       &BKF-I      &BKF-II     &IEKF\\
% \hline
%                    p=0.1   &10.949    &12.030     &11.469    &10.433\\
% \hline
%                    p=0.3   &16.655    &20.670     &23.407  &18.029\\
% \hline
%                    p=0.5   &39.794     &45.622     &73.435   &42.465\\
% \hline
%                    p=0.7   &87.039     &129.94     &131.76    &124.90\\
% \hline
% \end{tabular}
%\end{table}
\begin{figure}[!t]
	\centering
	\includegraphics[width=\linewidth]{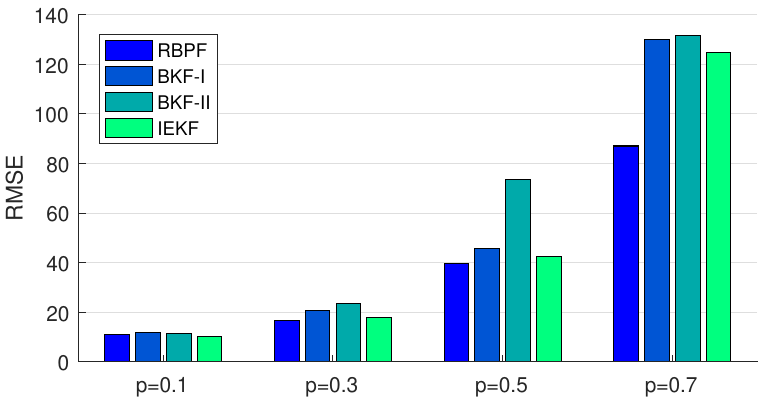}
	\caption{The RMSE of position under different levels of random measurement losses.}
	\label{fig1:exp1}
\end{figure}
 To track the system, we use the proposed three filters and the IEKF, which requires to know the measurement loss, and run $1500$ independent experiments for each filter to compute the RMSE of the position per time step.  The RBPF adopts $200$ particles. 

\begin{figure}[!t]
	\centering
	\includegraphics[width=\linewidth]{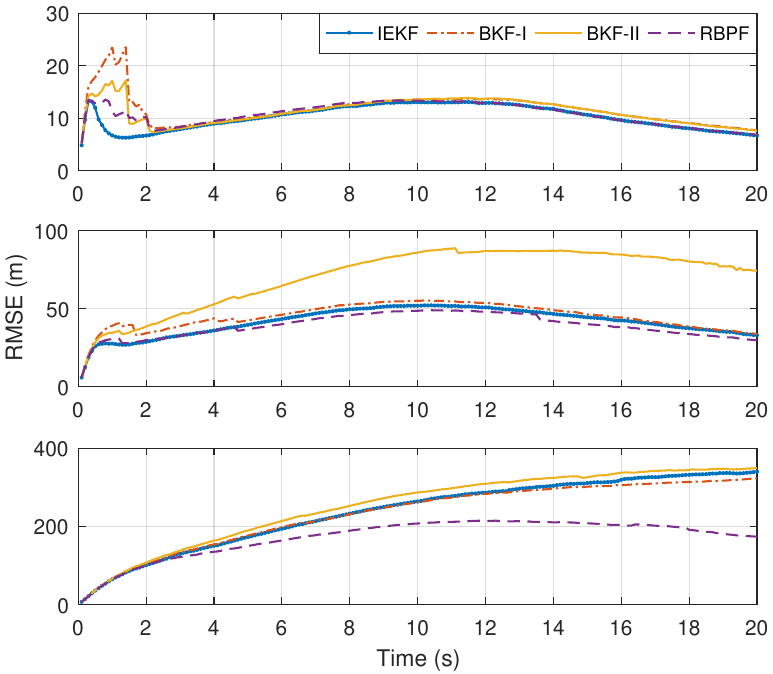}
	\caption{The RMSE of position under different levels of random measurement losses, e.g., $p=0.1$, $p=0.5$ and $p=0.8$.}
	\label{fig2:exp1}
\end{figure}

From Fig. \ref{fig1:exp1} and Fig. \ref{fig2:exp1}, one can observe that the RMSE of the RBPF is generally smaller than that of the BKF-I and BKF-II, and the BKF-I appears to outperform the BKF-II in this example. However, we cannot conclude that the BKF-I is better than the BKF-II. Note that the computational complexities of these two filters come close to each other.  Clearly, the larger the probability of random measurement losses, the larger the RMSE as expected.
%\begin{figure}[!t]
%\centering
%\captionsetup{justification=raggedright}
%\subfloat[]{\label{fig1a:exp1}\includegraphics[width=0.5\linewidth]{figs/p0dot1error.eps}}
%\subfloat[]{\label{fig1b:exp1}\includegraphics[width=0.5\linewidth]{figs/p0dot3error.eps}}\
%\subfloat[]{\label{fig1c:exp1}\includegraphics[width=0.5\linewidth]{figs/p0dot5error.eps}}
%\subfloat[]{\label{fig1d:exp1}\includegraphics[width=0.5\linewidth]{figs/p0dot7error.eps}}
%\caption{RMSE of position over 1500 times under different $p$. (a) $p=0.1$. (b) $p=0.3$. (c) $p=0.5$. (d) $p=0.7$.}
%\label{fig1:exp1}
%\end{figure}

It is interesting to observe that the estimation accuracy of the RBPF is better than that of the IEKF in many cases, especially when the probability of measurement losses $p$ is large. It is known that IKF is optimal when both $f$ and $h$ in \eqref{nonlinear} are linear functions and $\gamma_k$ is known. Under this case, the proposed filter cannot perform better than the IKF. However, the IEKF is not optimal if either $f$ or $h$ is nonlinear. In such cases, the RBPF may achieve a better performance than the IEKF due to the non-linearity in the measurement equation (\ref{tracking}).

\begin{figure*}[!t]
	\centering
	\subfloat[]{\label{fig3a:exp1}\includegraphics[width=0.33\linewidth]{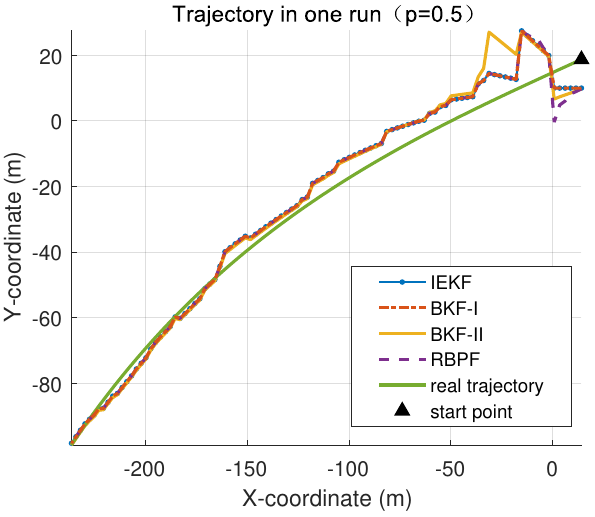}}
	\subfloat[]{\label{fig3b:exp1}\includegraphics[width=0.33\linewidth]{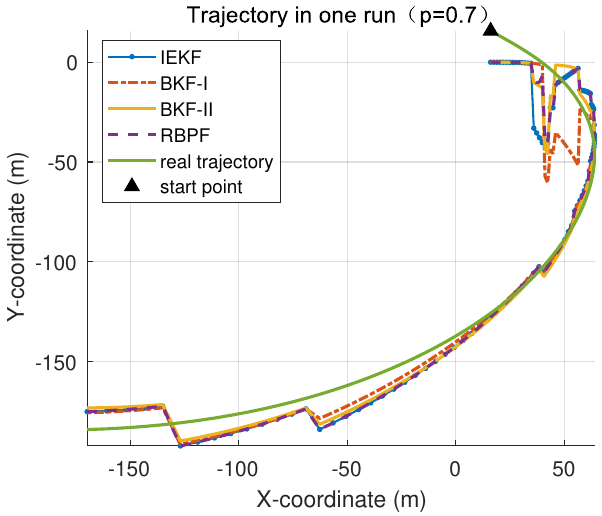}}
	\subfloat[]{\label{fig3c:exp1}\includegraphics[width=0.33\linewidth]{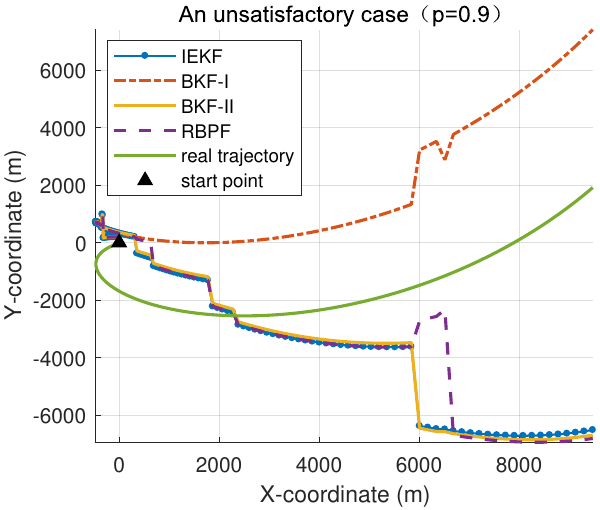}}
	\caption{One run trajectory.}
	\label{fig3:exp1}
\end{figure*}

Fig. \ref{fig3a:exp1} and Fig. \ref{fig3b:exp1} show the true and estimated trajectories in one experiment when $p=0.5$ and $p=0.7$ respectively. We can observe that our filters track the target well, especially under the low probability of measurement losses.

Fig. \ref{fig3c:exp1} indicates that the proposed filters perform unsatisfactorily under $p=0.9$ and the poor initial estimate, i.e., $\hat{x}_{0|0}^{(i)}=[200, 0, 0]^\text{T},\ i\in\{1,2\}$. In this case, all the filters track the trajectory badly, including the IEKF. This is due to the non-linearity of the measurement equation and the poor initial state estimate.  Once the filters have an inaccurate estimate at time $k$, the linearization step will further introduce estimation errors  and tends to a bad estimate. Similar to the standard EKF, a good initial estimate is essential for nonlinear systems. 

To summarize, the performance of the proposed filters varies from different systems because of the non-linearity of the systems. One can not conclude that one filter always performs better than another. Usually, the RBPF outperforms the other filters by increasing the number of particles.

\subsection{Quadrotor's Path Control}
\label{sec_quadrotor}
\begin{figure}[!t]
	\centering
	\includegraphics[width=6cm]{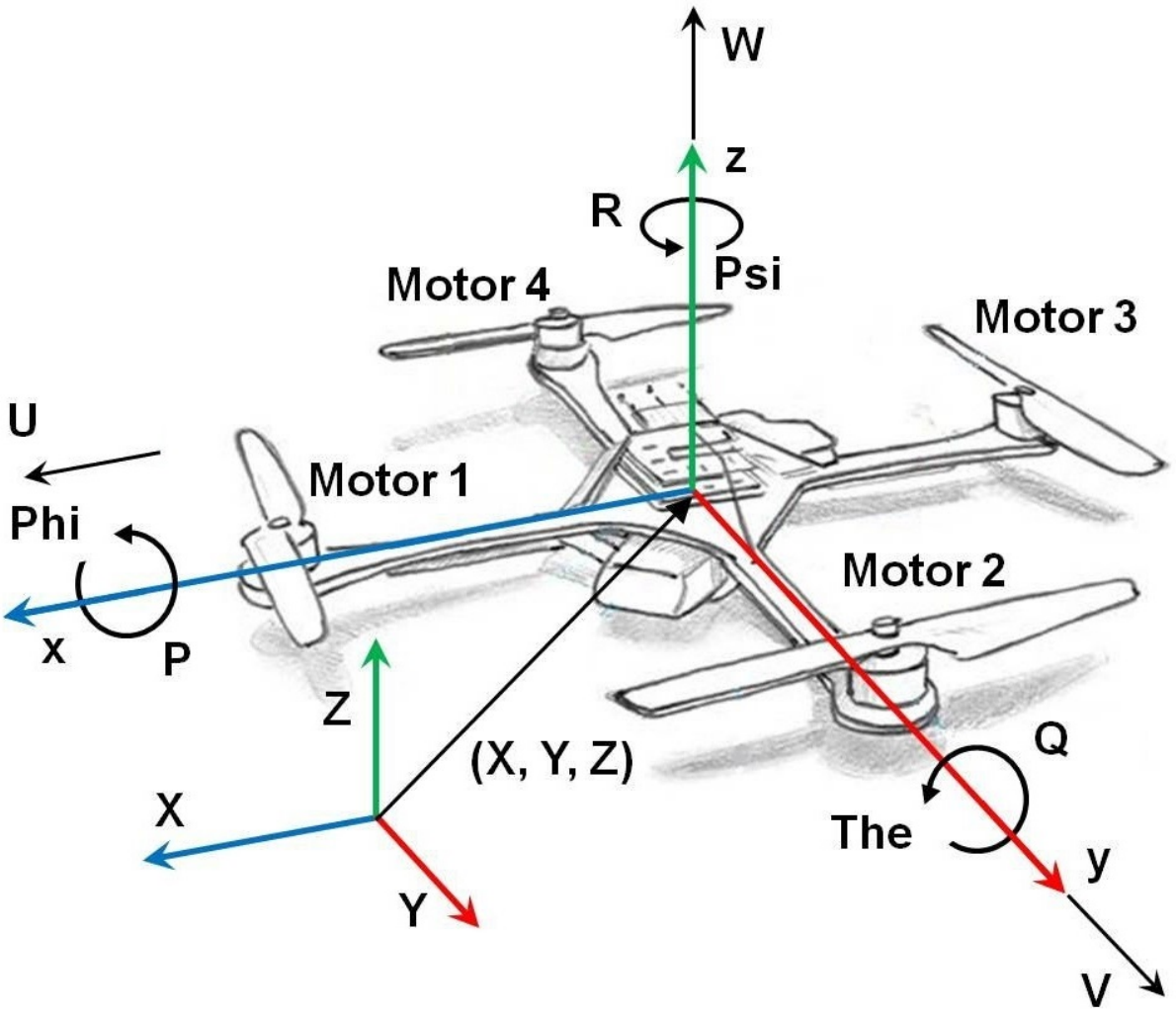}
	\caption{The structure of a quadrotor \cite{QuadSim}.}\label{fig0a:exp2}
\end{figure}

Finally, we test our filters on a quadrotor's path control problem where the state estimate is used for the feedback design.  A quadrotor is a very maneuverable unmanned aircraft with four rotors, and has been popular both in research and real applications over last years. The structure of quadrotors is shown in Fig. \ref{fig0a:exp2}  \cite{QuadSim}, where $[X,Y,Z]$ and $[\phi,\theta,\psi]$ are position and angular of the quadrotor in the inertia frame, respectively. $[U,V,W]$ and $[P,Q,R]$ are position and angular velocities in the body frame. Due to page limitation, we omit the dynamical model of quadrotors, see \cite{kun2015linear,crassidis2016three,dougherty1994use,mahony2012multirotor,bresciani2008modelling} for more details.

\begin{figure}[!t]
	\centering
	\includegraphics[width=7cm]{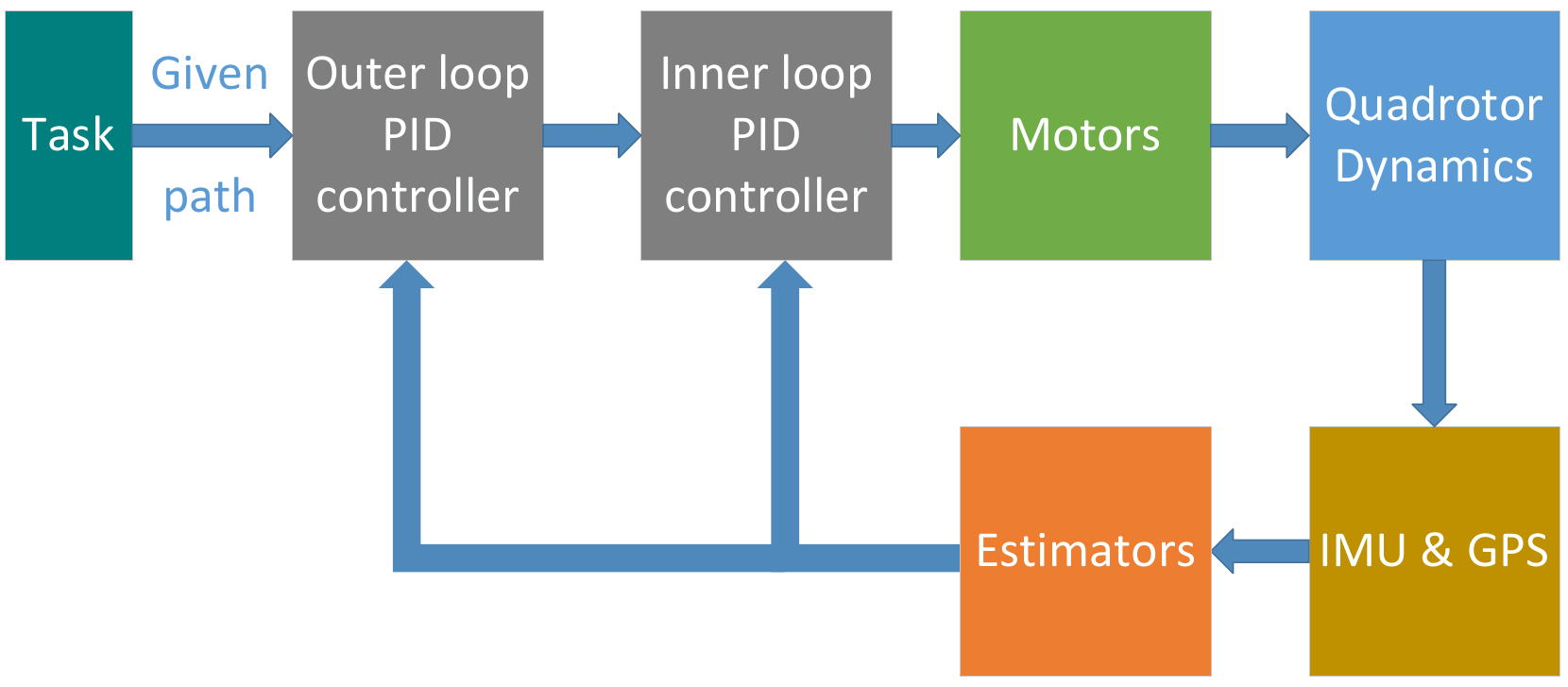}
	\caption{The control architecture of the quadrotor's path control system.}\label{fig0b:exp2}
\end{figure}

The control architecture is illustrated in Fig. \ref{fig0b:exp2}. The IMU and GPS block provides the noisy measurements of $[\phi,\theta,\psi]$, $[P,Q,R]$ and $[X,Y,Z]$, $[U,V,W]$ respectively. Due to bad GPS data and/or the unstable IMU, an estimator may only receive the true measurements intermittently. To this end, the proposed filters are applied to this situation and the estimated state are further sent to the outer and inner loop PID controller to control both the quadrotor's position and attitude respectively. This is realized by building a Matlab Simulink model based on a Matlab package called Quadcopter Dynamic Modeling and Simulation \cite{QuadSim}.

\begin{figure}[!t]
	\centering
	\includegraphics[width=\linewidth]{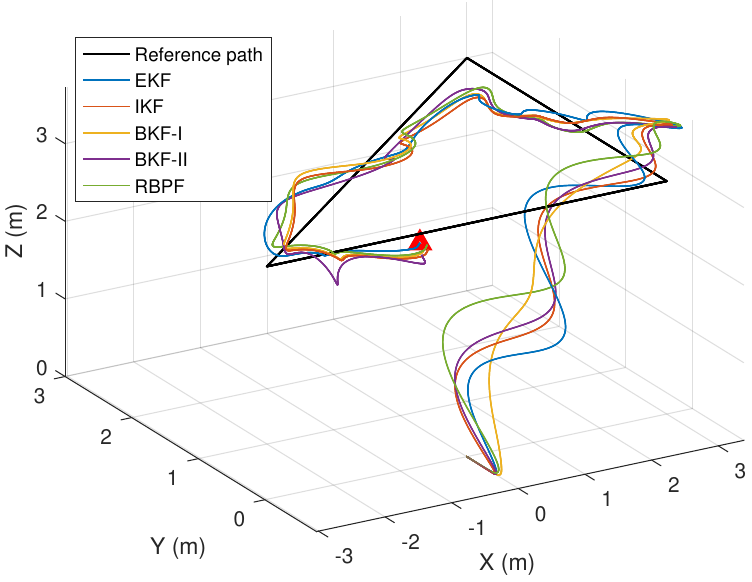}
	\caption{The trajectory of the controlled quadrotor in space under different filters. The red triangle mark is the end point of the trajectories.}
	\label{fig2:exp2}
\end{figure}

\begin{figure}[!t]
	\centering
	\includegraphics[width=\linewidth]{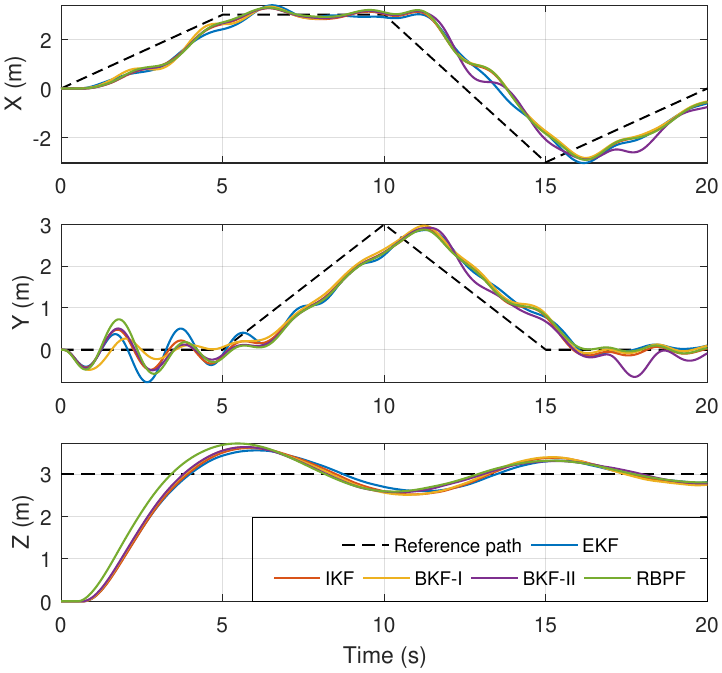}
	\caption{The trajectory of the controlled quadrotor under different filters in X, Y, and Z directions.}
	\label{fig1:exp2}
\end{figure}

Our goal is to control the quadrotor to follow a given triangle path in space, see Fig. \ref{fig2:exp2} (black line), and the projected trajectories in X,Y and Z directions are shown in black dashed lines in Fig. \ref{fig1:exp2}. We consider that the IMU and GPS measurements are sent as a single packet, and they will be either simultaneously lost or received by the estimator. Moreover, the probability of measurement losses is set to be $p=0.2$. Parameters of the PID controller are selected to be the same as those in \cite{QuadSim}, which  however does not consider the problem of measurement losses, and are provided in Table \ref{tab:exp2}.

\begin{table}
	\centering
	\caption{Parameters in the quadrotor's path control}\label{tab:exp2}
	\renewcommand{\arraystretch}{1.3}
	\begin{tabular}{|c|c|c|}
		\hline
		Parameter                               & Description                     & Value                   \\
		\hline
		$m$                                     & Quadrotor mass                  & 1.023kg                 \\
		$J_{xx},J_{yy}$                         & Moments of inertia              & $9.5$g$\cdot\text{m}^2$ \\
		$\sigma_\phi,\sigma_\theta,\sigma_\psi$ & \begin{tabular}[c]{@{}c@{}}Standard deviation of\\angular measurement noise\end{tabular}      & $0.1^\circ$             \\
		$\sigma_X,\sigma_Y,\sigma_Z$            & \begin{tabular}[c]{@{}c@{}}Standard deviation of\\position measurement noise\end{tabular}      & $2$m                    \\
		$T$                                     & Control frequency               & 100Hz                   \\
		$p$                                     & Measurements' error probability & 0.2                     \\
		\hline
	\end{tabular}
\end{table}

The trajectories of the closed-loop system are shown in Fig. \ref{fig2:exp2} and Fig. \ref{fig1:exp2}, where the reference path represents the desired trajectory. The trajectory of `EKF' denotes the controlled path of using the standard EKF without any measurement loss, and `IEKF' is that of using the IEKF to estimate the state, which relies on the true value of $\gamma_k$ at each time step. This is different from `BKF-I', `BKF-II', and `RBPF' where the packet loss process $\{\gamma_k\}$ is unknown to the estimator.
For comparison, all filters are tested under the same input noise, measurement loss process $\{\gamma_k\}$, initial conditions and covariance matrices.

\begin{figure}[!t]
	\centering
	\subfloat[]{\label{fig2a:exp2}\includegraphics[width=\linewidth]{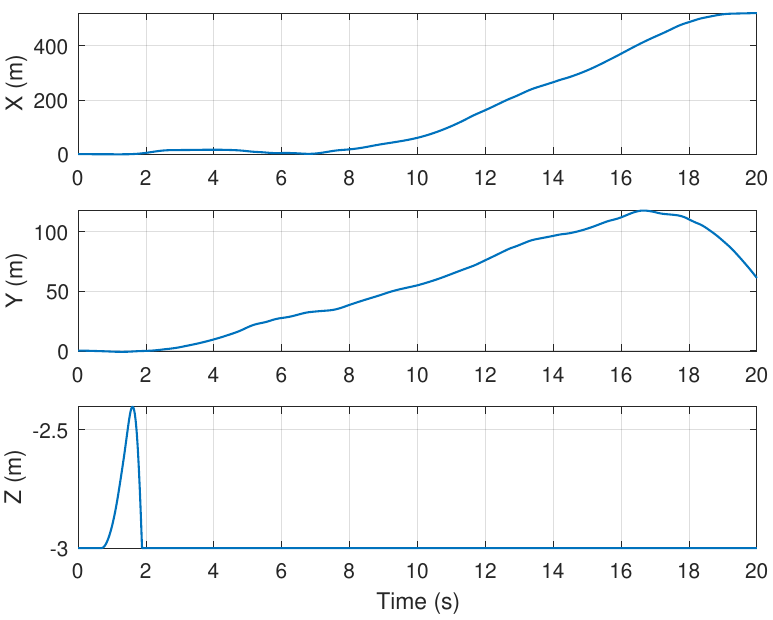}}
	%\subfloat[]{\label{fig2b:exp2}\includegraphics[width=0.5\linewidth]{figs/ex2obser.eps}}
	\caption{Tracking error of the EKF without considering the measurement losses.}
	\label{fig3:exp2}
\end{figure}

In view of Fig. \ref{fig2:exp2} and Fig. \ref{fig1:exp2}, it is not difficult to observe that all the proposed filters can fulfill the tracking task and their performance comes close to that of the standard EKF without any measurement loss. In contrast, we are unable to control the quadrotor to follow the desired trajectory if we directly apply the EKF update to all time steps without considering the measurement losses, see Fig. \ref{fig3:exp2}, where the tracking error diverges.

\section{CONCLUSION}
\label{sec_conclusion}
In this paper, we have designed three suboptimal filters to deal with the state estimation problem with unknown measurement losses. Among these filters, the BKF-I and the BKF-II were established by using the Bayesian point of view and the IEKF. Some approximations were made to obtain recursive forms. The RBPF is a particle filter based numerical method with a relatively small number of particles. All these proposed filters were applied to three different application problems, showing their effectiveness. Future work will focus on the theoretically convergence of the proposed filters.

\bibliographystyle{IEEEtran}
\bibliography{mybibf}         % bib file to produce the bibliography
\begin{IEEEbiography}
%[{\includegraphics[width=1in,height=1.25in,clip,keepaspectratio]{zhang.pdf}}]
{Jiaqi Zhang} received the B.S. degree in electronic and information engineering from Beijing Jiaotong University, Beijing, China, in 2016. He is currently pursuing the Ph.D. degree at the Department of Automation, Tsinghua University, Beijing, China. His research interests include networked control systems, distributed optimization and their applications.
\end{IEEEbiography}

\begin{IEEEbiography}
%[{\includegraphics[width=1in,height=1.25in,clip,keepaspectratio]{you.pdf}}]
{Keyou You}   received the B.S. degree in statistical science from Sun Yat-sen University, Guangzhou, China, in 2007 and the Ph.D. degree in electrical and electronic engineering from Nanyang Technological University (NTU), Singapore, in 2012. After briefly working as a Research Fellow at NTU, he joined Tsinghua University, Beijing, China in 2012 where he is currently an Associate Professor with the Department of Automation. He held visiting positions at Politecnico di Torino, Turin, Italy, the Hong Kong University of Science and Technology, Hong Kong, and the University of Melbourne, Parkville, VIC, Australia. His current research interests include networked control systems, distributed optimization and learning, and their applications.

Dr. You was a recipient of the Guan Zhaozhi Award at the 29th Chinese Control Conference in 2010, the CSC-IBM China Faculty Award in 2014, and the National Science Fund for Excellent Young Scholars in 2017.
\end{IEEEbiography}
\begin{IEEEbiography}
%[{\includegraphics[width=1in,height=1.25in,clip,keepaspectratio]{xie}}]
{Lihua Xie} received the B.E. and M.E. degrees in electrical engineering from Nanjing University of Science and Technology in 1983 and 1986, respectively, and the Ph.D. degree in electrical engineering from the University of Newcastle, Australia, in 1992. Since 1992, he has been with the School of Electrical and Electronic Engineering, Nanyang Technological University, Singapore, where he is currently a professor and Director, Delta-NTU Corporate Laboratory for Cyber-Physical Systems. He served as the Head of Division of Control and Instrumentation from July 2011 to June 2014. He held teaching appointments in the
Department of Automatic Control, Nanjing University of Science and Technology from 1986 to 1989 and Changjiang Visiting Professorship with South China University of Technology from 2006 to 2011.

Dr. Xie's research interests include robust control and estimation, networked control systems, multi-agent networks, localization and unmanned systems. He is an Editor-in-Chief for Unmanned Systems and an Associate Editor for IEEE Transactions on Network Control Systems. He has served as an editor of IET Book Series in Control and an Associate Editor of a number of journals including IEEE Transactions on Automatic Control, Automatica, IEEE Transactions on Control Systems Technology, and IEEE Transactions on Circuits and Systems-II. He is an elected member of Board of Governors, IEEE Control System Society (Jan 2016-Dec 2018). Dr. Xie is a Fellow of IEEE and Fellow of IFAC.
\end{IEEEbiography}

\end{document}